\documentclass[journal]{IEEEtran}

\IEEEoverridecommandlockouts
\usepackage{pifont,cite}
\usepackage{amsmath,amssymb,amsfonts,comment}
\usepackage{algorithmic}
\usepackage{graphicx}
\usepackage{textcomp}
\usepackage{xcolor,balance}
\usepackage[ruled,vlined]{algorithm2e}
\def\BibTeX{{\rm B\kern-.05em{\sc i\kern-.025em b}\kern-.08em
  T\kern-.1667em\lower.7ex\hbox{E}\kern-.125emX}}

\newcommand{\tr}[1]{\mathrm{tr}\!\left[{#1}\right]}

\newcommand{\xmark}{\ding{55}}%

\begin{document}

\title{A Low-Complexity Design for IRS-Assisted Secure Dual-Function Radar-Communication System  
}
\author{{Yi-Kai Li, \IEEEmembership{Member, IEEE}} and {Athina Petropulu, \IEEEmembership{Fellow, IEEE}\vspace{-6mm}}

\thanks{

This work was presented in part in 2023 IEEE Asilomar Conference on Signals, Systems, and Computers \cite{Li2023anIRS}. This work is supported by ARO grant 
W911NF2320103 and   
NSF  grant ECCS-2320568.

Athina Petropulu is with the Electrical and Computer Engineering Department, Rutgers, the State University of New Jersey, NJ, USA 08854 (e-mail: athinap@rutgers.edu).

Yi-Kai Li is with the Department of Electrical and Computer Engineering and Technology, Minnesota State University Mankato, MN, USA 56001 (e-mail: yikai.li@mnsu.edu).

}
}

\maketitle

\begin{abstract}
In dual-function radar-communication  (DFRC) systems the probing signal contains information intended for the communication users, which makes that information vulnerable to eavesdropping by the targets.  We study the security of a DFRC system aided by an intelligent reflecting surface (IRS) from the physical layer security (PLS) perspective. The IRS  helps overcome path loss
or blockage and introduces more degrees of freedom for system
design, however, it also makes the design problem more challenging.
%
In the system considered, the radar embeds artificial noise (AN)  in the probing waveform, and the 
  radar waveform, the AN noise and the IRS parameters are designed to  optimize the communication secrecy rate  while meeting radar signal-to-noise ratio (SNR) constraints. The contribution of the paper is a novel, low complexity approach to solve the underlying optimization problem and obtain the design parameters.
  In particular, we consider an alternating optimization approach, where in each iteration, the problem is decomposed into two sub-problems, namely, one that designs the IRS parameters, and another that  jointly designs the radar waveform and the AN. 
The  challenges in  those {sub-problems}  are the fractional objective, the  SNR being  a quartic function of the IRS parameters, and the unit-modulus constraint {on} the IRS parameters. A  fractional programming technique is used to transform the fractional form objective into a more tractable non-fractional polynomial form. A  closed-form {based} approach  is proposed for the IRS design problem, {which results in low complexity IRS design.} Numerical results are provided to demonstrate the convergence properties  of the proposed system design method,    
 the secrecy rate  and beamforming performance of the designed system.

\end{abstract}



\section{INTRODUCTION}

Integrated sensing and communication (ISAC) systems aim to perform both sensing and communication functions from a common platform \cite{Chen2020performance,Feng2020joint,Zhang2021anoverview,Sturm2011waveform}. 
\textcolor{black}{
An ISAC system  can sense the physical environment and also transmit communication signals, which greatly benefits 
unmanned aerial vehicles (UAV)  networks \cite{Mu2023UAV}, extended reality (XR) applications \cite{Ma2023integrated}, ultra-reliable and low-latency communications (URLLC) 
\cite{Zhao2022joint}, and autonomous vehicles \cite{Zhong2022empowering}.
}
A subclass of ISAC systems is the Dual-Function Radar-Communication (DFRC) systems, which, in addition to providing a common platform for both communication and sensing, utilize a shared waveform  \cite{Liu2020joint,Hassanien2019dual,Ma2020joint}.
%
In DFRC systems, user information is embedded in the probing waveform, resulting in higher spectral efficiency as compared to general ISAC systems. However, this also raises security concerns as the embedded communication information can  be intercepted by a target which is also an eavesdropper (ED).
{This paper studies the security of DFRC systems from a physical layer security (PLS) perspective.}

{In wireless communications, the  PLS design}
aims to leverage the physical characteristics of the wireless channel to degrade the quality of the signal received by the ED
{\cite{Wyner1975thewire,Fakoorian2011solutions,Dong2010improving,Zheng2011optimal,Li2011oncooperative,Zheng2013improving,Khisti2010secure,Goel2008guaranteeing}}. 
One approach to ensure PLS is  cooperative jamming, where 
trusted relays act  as helpers and  beamform artificial noise (AN), aiming to degrade the ED's channel \cite{Fakoorian2011solutions,Dong2010improving,Zheng2011optimal,Li2011oncooperative,Zheng2013improving}. Another approach is to embed AN in the signal intended for the users, in a way that it does not create interference to the users \cite{Khisti2010secure,Goel2008guaranteeing}.
PLS design for DFRC systems has been considered in \cite{Su2021secure,Su2023sensing}, 
 where the DFRC system  embeds AN in the transmit waveform, and a radar beamformer and the AN are jointly designed  to minimize the ED signal-to-noise-ratio (SNR) \cite{Su2021secure}, or maximize the weighted normalized {F}isher information matrix determinant and normalized secrecy rate \cite{Su2023sensing}.

 {A DFRC system} can be enhanced via the use of  intelligent reflecting surfaces (IRS).  
The IRS  is a planar array  of passive elements, each of which can alter the phase of the incoming electromagnetic wave in a computer-controlled manner. These elements can cooperatively perform beamforming to increase the power  that is radiated towards intended directions, or decrease it towards unintended directions. 
{The}
IRS can assist the DFRC system in overcoming performance limitations caused by path loss or blockage, which are likely to arise in the next-generation wireless systems due to the use of high frequencies.
{The} IRS can also create additional links between the radar and the users, or between the radar and the targets, thus introducing  more degrees of freedom (DoFs) for  system design \cite{Wei2022multiple,Jiang2021intelligent,Li2022dual,Li2023minorization,Liu2022joint}. When there is no  line-of-sight (LoS) between the radar and the target, which is also the case considered in this paper, the IRS can provide alternative paths for the radar signal to reach the target \cite{Wei2022multiple,Li2022dual}. 

 In this paper, we investigate the design for an IRS-aided secure DFRC system from the PLS perspective. The radar transmits a precoded waveform along with additive precoded AN.  {The
radar waveform, the AN noise and the IRS parameters are designed
to optimize the communication secrecy rate while meeting
radar SNR constraints.
 The IRS parameter is a diagonal matrix containing the responses of the IRS elements. We consider passive IRS elements, i.e., elements that only change the phase of the impinging signal.
 Therefore, each non-zero element of the IRS parameter matrix is subject to unit modulus constraint (UMC). 
Here, we consider the scenario where there is no LoS between the radar and the target.  The signal transmitted by the radar reaches the target after reflection {upon} the IRS, and subsequently returns to the radar {after reflection} upon the IRS \cite{Li2023intelligent}.
Thereby, the radar SNR is  a non-convex fourth order function of the IRS parameter. 
Since the design variables are highly coupled with each other, we solve the optimization problem via  an iterative optimization framework {where,} in each iteration, we first solve for the waveform and the AN by taking the IRS parameters as fixed (first sub-problem), and then {find} the IRS parameters 
with the previously 
 {computed} values for the waveform and the AN (second sub-problem).}
{The first sub-problem can be easily formulated as a quadratic programming problem.
On the other hand, the second sub-problem is more challenging 
 due to the non-convex
fractional form of the secrecy rate objective, the non-convex
high-order radar SNR, and the UMC on the IRS parameter.}

%

In our preliminary work \cite{Li2023anIRS}, we address the {aforementioned} 
second sub-problem
{via} a classical semidefinite relaxation (SDR) method. We  express  the non-convex high-order radar SNR term as a quadratic function  of an auxiliary variable, which is quadratic in the IRS parameter.  Subsequently,  via the  minorization maximization (MM) technique \cite{Sun2017majorization}, we  replace the SNR with a lower bound  that is linear in the auxiliary variable.
To address the non-convex fractional objective of secrecy rate and achieve a  tractable design problem we  invoke a fractional programming technique \cite{Shen2018fractional}. {Subsequently,} all functions in the objective and constraints are non-fractional quadratic functions of the IRS parameter, and the transformed problem is solved via an  SDR method. 
{The quadratic functions of the IRS parameter are transformed into functions of the 
{Gram}
 of the IRS parameter, 
and the {Gram} matrix is obtained via the interior point method (IPM).} However, in \cite{Li2023anIRS}
the UMC on the IRS parameter is relaxed during the optimization process, and the obtained solution is retracted to the unit complex circle to ensure feasibility. The retracted solution does not necessarily satisfy the radar SNR constraint, and at the same time leads to approximation loss. Thus, multiple  solutions need to be generated to obtain a satisfactory one.
{The complexity of the approach of \cite{Li2023anIRS}}
is $\mathcal O(N^{4.5})$. 
{When} large size IRS is desired to reap high array gain \cite{Najafi2021physics}, {such complexity becomes prohibitive.} 

To circumvent the IPM, we propose a novel closed-form expression-based IRS update method.
Firstly, the high order and fractional functions of the IRS parameter in the problem are transformed into linear functions. Specifically, the  radar SNR, is  converted {into}
a linear function of the IRS parameter  by applying {the} MM \cite{Sun2017majorization} twice \cite{Li2023efficient}. 
The  secrecy rate objective is 
{first} transformed into a quadratic non-fractional function via {the} quadratic transform \cite{Shen2018fractional}, {and} then it is  converted into a linear function {via the}  MM method \cite{Sun2017majorization}. As such, the IRS parameter design problem becomes a  linear programming (LP) problem subject to UMC. This problem can be solved by solving the corresponding Lagrangian dual problem, which has a closed-form solution in each iteration \cite{He2022qcqp}. 
The Lagrangian multiplier in the dual problem can be found by a bi-section search. \cite{He2022qcqp} introduced a method for non-fractional quadratic constrained quadratic programming (QCQP) under UMC, without incorporating high-order or fractional functions. \cite{He2022qcqp} also provided a case study on hybrid beamforming design utilizing their proposed solution. As compared to \cite{He2022qcqp} we propose a novel closed form expression by leveraging a solution that accommodates UMC in a different context. Unlike  \cite{Li2023anIRS}, in the proposed approach, the  UMC is always satisfied during the optimization process.

{\textbf{\underline{Relation to the literature}}
 The literature on striking a balance between PLS and sensing performance for IRS-aided DFRC systems is in its early stages. In \cite{Wang2023star,Hua2023secure,Chu2023joint,Zhang2023irs,Sweta2023efficient,Zhao2023joint,Salem2022active,Mishra2022optm3sec}, the waveform design problems are formulated as quadratic programming problems.} 
  In \cite{Wang2023star}, the waveform is designed by applying Courant’s penalty method, and the IRS is optimized by utilizing distance-majorization. In \cite{Hua2023secure}, a penalty-based algorithm is used in the waveform design problem, where auxiliary 
variables are defined, and
 the violation of the defined linear relationship between the original and auxiliary variables becomes a penalization term incorporated to the objective. The IRS design in \cite{Hua2023secure} is accomplished via MM. In \cite{Chu2023joint}, the  design of the waveform and IRS parameter is done via SDR and MM, respectively. In \cite{Zhang2023irs}, a penalization term is integrated to the objective to penalize the violation of  the  rank-one constraints of the {Gram} matrices of the design variables. In \cite{Sweta2023efficient}, the  design of IRS-aided secure DFRC system is completed by the MATLAB build-in \textit{fmincon} function. In \cite{Zhao2023joint}, the Dinkelbach’s transform \cite{Dinkelbach1967onnonlinear} is used to transform the waveform and IRS design problems to solvable non-fractional quadratic semi-definite programming (SDP) problems. In \cite{Salem2022active}, the waveform design problem is convexified by a fractional programming technique, and the IRS parameter is 
 {obtained} by applying the MM. In \cite{Mishra2022optm3sec}, the waveform design is convexified by a first-order Taylor series approximation, and the IRS is designed by a numerical simultaneous perturbation
stochastic approximation method.

However, as in  \cite{Li2023anIRS}, the IRS design in  most of the aforementioned methods
\cite{Chu2023joint,Zhang2023irs,Sweta2023efficient,Zhao2023joint,Salem2022active,Mishra2022optm3sec} are based on IPM or numerical methods, which are computationally intensive when the IRS size is large. Our paper proposes a closed-form expression based method to enable a scalable IRS design algorithm. 
{For those that also have a closed-form expression solution \cite{Wang2023star,Hua2023secure} for the IRS design,} \cite{Wang2023star} studies a different scenario, i.e., symbol-level precoding for secure IRS DFRC system. 
The metrics involved in the design problem are non-fractional quadratic functions. Our proposed method addresses optimization problem containing high order and fractional functions of the IRS parameters. In \cite{Hua2023secure}, the authors focus on the problem of maximizing the  beampattern gain towards the target subject to minimum SINR at the user and maximum allowed information leakage to the target/ED, which is a different problem as compared to our paper. Moreover, \cite{Hua2023secure} concentrates on the metric of beampattern gain towards the target without addressing secrecy rate, while our paper {involves} the secrecy rate at the user,  the SNR at radar receiver, and {the} obtained beampatterns under different system configurations. In addition to the aforementioned methods,  deep reinforcement learning (DRL) \cite{Evmorfos2023actor} have been used to optimize the secrecy rate of IRS-aided DFRC system \cite{Liu2023drl}. 
 However, the dimensionality of the agent's search space, specifically the configured phases of the IRS elements, expands exponentially as the size of the IRS increases. Consequently, DRL algorithms might exhibit a slower rate of convergence in scenarios where the IRS is  large.

\vspace{-0mm}
\section{SYSTEM MODEL}
\label{sec:models}

\begin{figure}[!t]\vspace{0mm} 
	\hspace{5mm} 
	\def\svgwidth{240pt} 
	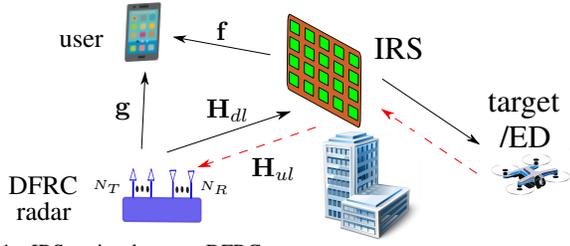 \vspace{-8mm}
	\caption{IRS-assisted secure DFRC system.}\vspace{-5mm} 
	\label{fig:system_model}
\end{figure}

We consider an IRS-aided DFRC system, as show in Fig. \ref{fig:system_model}, where the DFRC is communicating with a user 
{while} simultaneously tracking a non-line-of-sight (NLoS) target. {We consider such NLoS case as that case necessitates the use of an IRS for target sensing.}
{The target is not authorized to access the communication information, so it will be treated as  ED.}
The DFRC transmitter is  an $N_T$-antenna uniform linear array (ULA), and the {DFRC} receiver is a collocated $N_R$-antenna ULA array. The inter-antenna distance in both arrays is denoted by  $d$. {Both the user and the ED are equipped with a single antenna.}   An $N$-element IRS is deployed to aid both the sensing and communication functionalities. The channels are assumed flat fading and the channel coefficients  are assumed to be available. 
{In DFRC systems, the channels can be estimated by leveraging the sensing function  \cite{Jiang2023sensing,Silvia2023enhanced}.}
{Here we assume perfect channel knowledge, however, in simulations we demonstrate the effects of channel errors on the system performance.}


 The signal transmitted  by the DFRC radar {is the sum of the precoded signal intended for the communication user, $s$, and the precoded AN, $\mathbf n$, i.e.,}
\begin{eqnarray}
	\mathbf x = \mathbf w  s + \mathbf W_n  \mathbf n,
\end{eqnarray}
\noindent where $\mathbf w \in \mathbb C^{N_T \times 1}$ denotes the   precoder for {$s$ and $\mathbf W_n \in \mathbb C^{N_T \times N_T}$ the precoder for $\mathbf n$. Here,} $s$ is assumed to be zero-mean white, with unit variance; $\mathbf n \sim \mathcal{CN}(\mathbf 0_{N_T \times 1}, \sigma^2 \mathbf I_{N_T})$. The AN is used for jamming but also helps in the  detection of the target/ED.

 Since there is no LoS between the radar and the target,  the transmitted waveform arrives at the target via the DFRC-IRS-target path, and the target echo reaches the radar receiver via the target-IRS-DFRC path. The signal at the radar receiver is

\vspace{-3mm}
\begin{small}
\begin{eqnarray} 
   \mathbf y_R \!\!\!&=&\!\!\!
    {\alpha_T \beta \mathbf H_{ul} \mathbf \Phi \mathbf a_I(\psi_{a},\psi_{e}) \mathbf a_I^{T}(\psi_{a},\psi_{e}) \mathbf \Phi \mathbf  H_{dl} }
    (\mathbf w  s \!+\! \mathbf W_n  \mathbf n)  \!+\! \mathbf n_R \nonumber\\  \!\!\!\!\!\!\!\!\!\!\!\!&=& \!\!\!\!\mathbf C_{T} (\mathbf w  s + \mathbf W_n  \mathbf n) + \mathbf n_R,  \label{eqn:y_R}
\end{eqnarray}
\end{small}
\vspace{-3mm}

\noindent where $\alpha_T$ {represents} the radar target reflection {coefficient}; $\beta$ is  the combined effect of the attenuation of DFRC-IRS-target-IRS-DFRC path, the antenna gain,    and any additional  losses or gains; $\mathbf H_{ul}$ and $\mathbf H_{dl}$ are the normalized IRS-DFRC and DFRC-IRS channels, respectively; $\mathbf \Phi = \text{diag}([e^{j \phi_1},e^{j \phi_2},\cdots,e^{j \phi_N}])$ represents the IRS parameter  {matrix, with} $\phi_n \in [0,2 \pi)$ {denoting} the phase shift of the $n$-th {IRS} element; $\mathbf a_I(\psi_a,\psi_e)$ is the steering vector of {the} IRS, {with} $\psi_a$ and $\psi_e$ {denoting} the azimuth and elevation angles of the target relative to the IRS, respectively; $\mathbf n_R \sim \mathcal{CN}(\mathbf 0_{N_R \times 1}, \sigma_R^2 \mathbf I_{N_R})$ {represents} additive white Gaussian noise (AWGN) at the DFRC receiving array.

The SNR at the radar receiver for detecting the target is
\begin{eqnarray}
	\gamma_R= \tr{ \mathbf C_{T} \left[ \mathbf w \mathbf w^H + \mathbf W_n \mathbf W_n^H \right] \mathbf C_{T}^H }/ \sigma_R^2. \label{eqn:gamma_R}
\end{eqnarray}
The received signal at the communication user is 
\begin{eqnarray} \label{eqn:y_U}
	 y_{u} &=& (\mathbf g^T + \beta_H^{\frac{1}{2}} \mathbf f^T \mathbf \Phi \mathbf H_{dl} ) (\mathbf w  s + \mathbf W_n  \mathbf n) +  n_u \nonumber \\ &=& \mathbf c_u^T (\mathbf w  s + \mathbf W_n  \mathbf n) +  n_u,
\end{eqnarray}
\vspace{-3mm}

\noindent where $\mathbf g \in \mathbb C^{N_T \times 1}$ is the  channel between the user and DFRC transmitter; $\beta_H$ is the path-loss of the DFRC-IRS channel; $\mathbf f \in \mathbb C^{N \times 1}$ denotes the   user-IRS channel; $n_u \sim \mathcal {CN}(0,\sigma_u^2)$ is   AWGN  at the user.

The received signal at the target/ED is

\vspace{-3mm}
\begin{eqnarray} \label{eqn:y_TE}
	y_{te} &=& ( \beta^{\frac{1}{2}} \mathbf a_I^T(\psi_{a},\psi_{e}) \mathbf \Phi \mathbf H_{dl} ) (\mathbf w  s + \mathbf W_n  \mathbf n) +  n_{te} \nonumber \\ &=& \mathbf c_{te}^T (\mathbf w  s + \mathbf W_n  \mathbf n) +  n_{te},
\end{eqnarray}
\vspace{-3mm} 

\noindent where $n_{te} \sim \mathcal {CN}(0,\sigma_{te}^2)$ is   AWGN  at the target/ED.

\noindent The achievable rates at the user and target/ED are respectively
\begin{eqnarray}
\!\!\!\!\!\!\!\!\!\!\!\!\!\!\!	&&R_u = \log \left( 1+\text{SINR}_{u} \right) = \log \left( 1+\frac{| \mathbf c_u^T \mathbf w|^2 }{ || \mathbf c_u^T \mathbf W_n||^2 + \sigma_u^2 } \right), \label{eqn:R_u} \\ \!\!\!\!\!\!\!\!\!\!\!\!\!\!\!&& R_{te} = \log \left( 1+\text{SINR}_{te} \right) = \log \left( 1+\frac{| \mathbf c_{te}^T \mathbf w|^2 }{ || \mathbf c_{te}^T \mathbf W_n||^2 + \sigma_{te}^2 } \right). \label{eqn:R_te}
\end{eqnarray}

\section{SYSTEM DESIGN}
\label{sec:sys_design_quadratic}
We consider the  design of the radar information precoder, $\mathbf w$, the artificial noise precoder, $\mathbf W_n$, and the IRS parameter matrix, $\mathbf \Phi$,
as that of maximizing the secrecy rate,  defined as $(R_u-R_{te})$, while satisfying certain constraints, i.e.,
\begin{subequations} \label{eqn:orig_prob}
	\vspace{-0.1in}
	\begin{eqnarray} 
		\!\!\!\!\!(\mathbb P) \;\;\;\;\;\; \max_{\mathbf w, \mathbf W_n, \mathbf \Phi}&&	\;\; R_u - R_{te} \label{eqn:obj_sr}  \\ \!\!\!\!\!\mathrm{s.t.} && \;\; \tr{\mathbf w \mathbf w^H+\mathbf W_n \mathbf W_n^H}  \leq P_R \label{eqn:tot_power}\\\!\!\!\!\!&& \;\;  |\mathbf \Phi_{n,n}| = 1,\;\; \forall n = 1, \cdots, N\label{eqn:unit_modu}\\\!\!\!\!\!&& \;\;  \gamma_R  \geq \gamma_{R,th} \label{3d}    
		\label{eqn:radar_snr}
	\end{eqnarray} 
\end{subequations}

\noindent where (\ref{eqn:tot_power}) enforces that the total power of the {transmitted signal} stays below $P_R$; (\ref{eqn:unit_modu}) enforces unit modules  elements in $\mathbf \Phi$. This is because the IRS is composed of passive elements, which can only change the phase of the impinging signal; (\ref{eqn:radar_snr}) ensures that the radar SNR \textcolor{blue}{is} above threshold $\gamma_{R,th}$.

\textit{Remark} {In the following we will assume that the target/ED angle is known. However, it does not need to be known exactly, and only an initial estimate would suffice. As long as the target will fall within the beam that will be formulated by the proposed system, it will be detected   and its angle will be fine-tuned. The new angle will be used in the subsequent communication/sensing phase. In this way, the proposed DFRC will be able to track moving targets. }

The problem $(\mathbb P)$   in (\ref{eqn:orig_prob}) is challenging for the following reasons: (i) The design variables,  $\mathbf w, \mathbf W_n, \mathbf \Phi$, are mutually coupled. (ii) The objective is the difference between two non-convex functions ($R_u$ and $R_{te}$), and is non-convex. (iii) The radar SNR in  (\ref{eqn:radar_snr}), $\gamma_R$, is quartic in $\mathbf \Phi$. Since $\gamma_R$ is quadratic in $\mathbf C_{T}$ {(see (\ref{eqn:gamma_R})),}  and $\mathbf C_{T}$ is second order in $\mathbf \Phi$ {(see (\ref{eqn:y_R})).} (iv) $\mathbf \Phi$ is subject to highly non-convex UMC. 
    
To decouple the design variables, we divide the problem (\ref{eqn:orig_prob}) into two sub-problems. The first one is optimization with respect to  $\mathbf w$ and $\mathbf W_n$ for fixed  $\mathbf \Phi$, and the second one is optimization with respect to $\mathbf \Phi$ for fixed  $\mathbf w$ and $\mathbf W_n$. These two sub-problems are solved in an alternating way until a convergence condition is met \cite{Li2017_coexistence}.

\subsection{First sub-problem: Solve for $\mathbf w$ and $\mathbf W_n$ for fixed  $\mathbf \Phi$} \label{sec:waveform_AN_design}

The first sub-problem is formulated as follows:

\vspace{-5mm}
\begin{subequations} \label{eqn:problem1}
	\begin{eqnarray} 
	\!\!\!\!\!\!\!\!\!(\mathbb P_{1})	\qquad \max_{\mathbf w, \mathbf W_n }&&	\;\;\;\;	R_u - R_{te}\label{eqn:obj1}\\ \!\!\!\!\!\!\!\!\!\mathrm{s.t.} && \;\;\;\; \tr{\mathbf w \mathbf w^H + \mathbf W_n \mathbf W_n^H}  \leq P_R \label{eqn:tot_power1}\\\!\!\!\!\!\!\!\!\!&& \;\;\;\; \gamma_R  \geq \gamma_{R,th} \label{eqn:radar_snr_cons1}
	\end{eqnarray}
\end{subequations}
\vspace{-5mm}

\noindent Note that $R_u$, $R_{te}$ are both 
fractional forms of the design variables. Here, we invoke the quadratic transform technique \cite{Shen2018fractional} to recast $R_u$ and $R_{te}$ into non-fractional functions of $\mathbf w$ and $\mathbf W_n$.  Thereby, the problem $(\mathbb P_{1})$ becomes a non-fractional quadratic programming problem. The Lagrangian dual expressions of $R_u$ and $R_{te}$ in (\ref{eqn:R_u}) and (\ref{eqn:R_te}) are 
\begin{eqnarray}
	 \!\!\!\!\!\!\!\!\!\!\!&&R_u 
	  =  {\left( 1+\gamma_{u} \right)| \mathbf c_u^T \mathbf w|^2 }/\left({ | \mathbf c_u^T \mathbf w|^2 + || \mathbf c_u^T \mathbf W_n||^2 + \sigma_u^2 }\right) - \gamma_{u}  \nonumber \\ \!\!\!\!\!\!\!\!\!\!\!\!\!\!\! && + \log \left( 1+\gamma_{u} \right) =  - \alpha_u^2 (| \mathbf c_u^T \mathbf w|^2+|| \mathbf c_u^T \mathbf W_n||^2 + \sigma_u^2) \nonumber \\ \!\!\!\!\!\!\!\!\!\!\!\!\!\!\!&& + 2 \alpha_u \sqrt{1+\gamma_u} |\mathbf c_u^T \mathbf w| + \log \left( 1+\gamma_{u} \right) - \gamma_{u},\label{eqn:Ru_dual} \\
	   \!\!\!\!\!\!\!\!\!\!\!\!\!\!\!&&R_{te} 
	  =  {\left( 1+\gamma_{te} \right)| \mathbf c_{te}^T \mathbf w|^2 } \!/\! \left({ | \mathbf c_{te}^T \mathbf w|^2 \!+\!|| \mathbf c_{te}^T \mathbf W_n||^2 \!+\! \sigma_{te}^2 }\right)  - \gamma_{te} \nonumber \\ \!\!\!\!\!\!\!\!\!\!\!\!\!\!\! && + \log \left( 1+\gamma_{te} \right) =  - \alpha_{te}^2 (| \mathbf c_{te}^T \mathbf w|^2 + || \mathbf c_{te}^T \mathbf W_n||^2 + \sigma_{te}^2) \nonumber \\ \!\!\!\!\!\!\!\!\!\!\!\!\!\!\!&&+2 \alpha_{te} \sqrt{1+\gamma_{te}} |\mathbf c_{te}^T \mathbf w|  + \log \left( 1+\gamma_{te} \right) - \gamma_{te}, \label{eqn:Rte_dual} 
\end{eqnarray}
 \noindent where $\gamma_{u}$, $\alpha_u$,  $\gamma_{te}$, and $\alpha_{te}$ are auxiliary variables, whose values are updated in each iteration. 
 The derivation details  of \eqref{eqn:Ru_dual} and \eqref{eqn:Rte_dual} are provided in   Appendix \eqref{sec:Lagrangian_dual}.
 Then, the objective can be re-written as 
 \begin{eqnarray}
 	R_u - R_{te} = c + \Re{(\mathbf v^T \mathbf w)} + \tr{\mathbf M (\mathbf W_n \mathbf W_n^H+\mathbf w \mathbf w^H)},
 \end{eqnarray}
\noindent where 
\begin{eqnarray}
	\!\!\!\!\!\!\!\!\!\!\!\!\!\!\!\!\!\!\!\!\!&&c = \log \left( 1\!+\!\gamma_{u} \right) \!-\! \gamma_u \!-\! \log \left( 1\!+\!\gamma_{te} \right) \!+\!\gamma_{te}\! +\! \alpha_{te}^2 \sigma_{te}^2 \!-\! \alpha_{u}^2 \sigma_{u}^2, \\\!\!\!\!\!\!\!\!\!\!\!\!\!\!\!\!\!\!\!\!\!&&
	\mathbf v = 2 \alpha_u \sqrt{1\!+\!\gamma_u} \mathbf c_u - 2 \alpha_{te} \sqrt{1\!+\!\gamma_{te}} \mathbf c_{te}, \label{eqn:v_vector} \\\!\!\!\!\!\!\!\!\!\!\!\!\!\!\!\!\!\!\!\!\!&& \mathbf M = \alpha_{te}^2  \mathbf c_{te}^{\ast} \mathbf c_{te}^{T} - \alpha_{u}^2  \mathbf c_{u}^{\ast} \mathbf c_{u}^{T}. \label{eqn:M_matrix}
\end{eqnarray}
 
\noindent The above is  a quadratic programming problem with non-fractional objective and constraints and can be solved via  SDR. By letting $\mathbf R_w =  \mathbf w^H \mathbf w$ \footnote{This constraint can not be directly accepted in CVX toolbox \cite{cvx} for $N > 1$, and is relaxed to an inequality as (\ref{eqn:Rw_ineq}).} and $\mathbf R_{W_n} =  \mathbf W_n^H \mathbf W_n$, the Problem  $(\mathbb P_{1})$ in (\ref{eqn:problem1}) becomes
\begin{subequations} \label{eqn:problem1_new}
	\begin{eqnarray} 
		\!\!\!\!\!\!\!\!\!\!\!(\overline{\mathbb P}_1) \!\!\!\!	\quad \max_{\mathbf w, \mathbf R_w \mathbf R_{W_n} }&&\!\!\!\!\!\!\!\! \Re{(\mathbf v^T \mathbf w)} + \tr{\mathbf M (\mathbf R_{W_n} + \mathbf R_{w})} + c  \label{eqn:obj1_new}\\ \!\!\!\!\!\!\!\!\!\!\!\mathrm{s.t.} && \!\!\!\!\!\!\!\! \tr{\mathbf R_w + \mathbf R_{W_n}}  \leq P_R \label{eqn:tot_power1_new}\\\!\!\!\!\!\!\!\!\!\!\!&& \!\!\!\!\!\!\!\! \tr{ \mathbf C_{T}^H \mathbf C_{T} \left[ \mathbf R_{w} + \mathbf R_{W_n} \right]  }/ \sigma_R^2  \geq \gamma_{R,th} \label{eqn:radar_snr_cons1_new} \\\!\!\!\!\!\!\!\!\!\!\!&& \!\!\!\!\!\!\!\! \mathbf R_w \succcurlyeq  \mathbf w \mathbf w^H \label{eqn:Rw_ineq}
	\end{eqnarray}
\end{subequations}
\noindent where $c$  is irrelevant to the design variables, and thus can be dropped, and $\mathbf R_w$ and $\mathbf R_{W_n}$ are  positive semidefinite matrices. The Problem $(\overline{\mathbb P}_1)$ in (\ref{eqn:problem1_new}) is an LP problem, whose optimal solution, say $\mathbf w^{\ast}$ and $\mathbf R_{W_n}^{\ast}$, can be obtained by numerical solvers, for example by using the CVX toolbox \cite{cvx}. The optimal solution for $\mathbf W_n$ is obtained by calculating the square root matrix of $\mathbf R_{W_n}^{\ast}$.

\subsection{Second sub-problem: Solve for $\mathbf \Phi$ for fixed $\mathbf w$ and $\mathbf W_n$}\label{sec:problem2_qtSDR}
The design problem of IRS parameter, $\mathbf \Phi$,  is as follows

\vspace{-5mm}
\begin{subequations} \label{eqn:problem2}
	\begin{eqnarray} 
	({\mathbb P}_2) \qquad	\max_{\mathbf \Phi}&&	\;\;\;\; R_u - R_{te}	\label{eqn:obj2}\\ \mathrm{s.t.} && \;\;\;\; 
		|\mathbf \Phi_{n,n}| = 1,\;\; \forall n = 1, \cdots, N\label{eqn:unit_modu2} \\ && \;\;\;\; \gamma_R \geq \gamma_{R,th} \label{eqn:radar_snr2}
	\end{eqnarray}
\end{subequations}
Besides the fractional form objective, the problem $({\mathbb P}_2)$ in (\ref{eqn:problem2}) is subject to the challenging non-convex UMC on $\mathbf \Phi$ as shown in (\ref{eqn:unit_modu2}). Moreover, the $\gamma_R$ in (\ref{eqn:radar_snr2}) is a non-convex fourth order function of $\mathbf \Phi$. This term needs to be convexified to make the problem solvable. 

We rewrite the effective channel for the user as
\begin{eqnarray}
	\!\!\!\!\!\!\!\!&&\mathbf c_u^T = 	\mathbf g^T + \beta_H^{\frac{1}{2}} \mathbf f^T \mathbf \Phi \mathbf H_{dl} =  \mathbf g^T + \boldsymbol{\phi}^T \beta_H^{\frac{1}{2}} \text{diag} (\mathbf f)  \mathbf H_{dl} \nonumber\\ \!\!\!\!\!\!\!\!&&=\mathbf g^T + \boldsymbol{\phi}^T \mathbf D, \label{eqn:c_u_T}
\end{eqnarray}
\noindent where $\boldsymbol{\phi} = \text{diag}(\mathbf \Phi)$ is the column vector that contains all diagonal elements of $\mathbf \Phi$, and $\mathbf D = \beta_H^{\frac{1}{2}} \text{diag} (\mathbf f)  \mathbf H_{dl}$. Similarly, the effective channel for the ED/target is re-written as
\begin{eqnarray}
	\!\!\!\!\!\!\!\!&&\mathbf c_{te}^T = \beta^{\frac{1}{2}} \mathbf a_I^T(\psi_{a},\psi_{e}) \mathbf \Phi \mathbf H_{dl} = \boldsymbol{\phi}^T \beta^{\frac{1}{2}} \text{diag} (\mathbf a_I(\psi_{a},\psi_{e}))  \mathbf H_{dl}\nonumber \\ \!\!\!\!\!\!\!\!&&= \boldsymbol{\phi}^T \mathbf E, \label{eqn:c_te_T}
\end{eqnarray}
\noindent where $\mathbf E=\beta^{\frac{1}{2}} \text{diag} (\mathbf a_I(\psi_{a},\psi_{e}))  \mathbf H_{dl}$. By invoking (\ref{eqn:v_vector}), (\ref{eqn:c_u_T}), and (\ref{eqn:c_te_T}), the first term in the transformed non-fractional objective in (\ref{eqn:problem1_new}), $\Re{(\mathbf v^T \mathbf w)}$, can be re-written as 
\begin{eqnarray}
	\Re{(\mathbf v^T \mathbf w)} = \Re{(\boldsymbol{\phi}^T \boldsymbol \eta)} + c_1,
\end{eqnarray}
\noindent where 
\begin{eqnarray}
\boldsymbol{\eta}&=&2(\alpha_u \sqrt{1+\gamma_u} \mathbf D - \alpha_{te} \sqrt{1+\gamma_{te}} \mathbf E) \mathbf w,\\ c_1 &=& 2 \Re{(\alpha_u \sqrt{1+\gamma_u}} \mathbf g^T \mathbf w).
\end{eqnarray}
Furthermore, $c_1$ is taken as a constant in the second sub-problem. Likewise, by referring to (\ref{eqn:M_matrix}),  (\ref{eqn:c_u_T}), and (\ref{eqn:c_te_T}),  the second item of the converted objective in (\ref{eqn:problem1_new}), $\tr{\mathbf M \mathbf R_{W_n}}$, is re-expressed as
\begin{eqnarray}
    &&\tr{\mathbf M \mathbf R_{W_n}} = \alpha_{te}^2  \mathbf c_{te}^{T} \mathbf W_n \mathbf W_n^H \mathbf c_{te}^{\ast} - \alpha_{u}^2  \mathbf c_{u}^{T} \mathbf W_n \mathbf W_n^H \mathbf c_{u}^{\ast} \nonumber \\ &&= \boldsymbol \phi^T \mathbf L_1 \boldsymbol \phi^{\ast} -\Re{(\boldsymbol{\phi}^T \boldsymbol \mu)} - c_2,
\end{eqnarray}
\noindent where 
\begin{eqnarray}
    \mathbf L_1 &=& \alpha_{te}^2  \mathbf E \mathbf W_n \mathbf W_n^H \mathbf E^H - \alpha_{u}^2  \mathbf D \mathbf W_n \mathbf W_n^H \mathbf D^H,\\ \boldsymbol \mu &=&2 \alpha_{u}^2  \mathbf D \mathbf W_n \mathbf W_n^H \mathbf g^{\ast}, \\ c_2 &=& \alpha_{u}^2  \mathbf g^T \mathbf W_n \mathbf W_n^H \mathbf g^{\ast}.
\end{eqnarray}
Likewise, $c_2$ is taken as a constant in the design of $\mathbf \Phi$ or 
$\boldsymbol{\phi}$. Similarly, $\tr{\mathbf M \mathbf R_{w}}$ can be re-written as
\begin{eqnarray}
    &&\tr{\mathbf M \mathbf R_{w}} = \boldsymbol \phi^T \overline{\mathbf L}_1 \boldsymbol \phi^{\ast} -\Re{(\boldsymbol{{\phi}^T \overline{\boldsymbol \mu})}} - \bar c_2,
\end{eqnarray}
\noindent where 
\begin{eqnarray}
    \overline{\mathbf L}_1 &=& \alpha_{te}^2  \mathbf E \mathbf w \mathbf w^H \mathbf E^H - \alpha_{u}^2  \mathbf D \mathbf w \mathbf w^H \mathbf D^H,\\ \overline{\boldsymbol \mu} &=&2 \alpha_{u}^2  \mathbf D \mathbf w \mathbf w^H \mathbf g^{\ast}, \\ \bar c_2 &=& \alpha_{u}^2  \mathbf g^T \mathbf w \mathbf w^H \mathbf g^{\ast}.
\end{eqnarray}
Thereby, the objective, $R_u - R_{te}$ can be re-written as
\begin{eqnarray}
\!\!\!\!\!\!\!\!\!\!\!\!	&&R_u - R_{te} = \Re{(\mathbf v^T \mathbf w)} + \tr{\mathbf M (\mathbf R_{W_n} + \mathbf R_{w})} + c \nonumber \\\!\!\!\!\!\!\!\!\!\!\!\!&&= \boldsymbol \phi^T (\mathbf L_1 +\overline{\mathbf L}_1)\boldsymbol \phi^{\ast} +\Re{[\boldsymbol{\phi}^T (\boldsymbol \eta-\boldsymbol \mu - \overline{\boldsymbol \mu})]} + \text{const}, \label{eqn:obj2_new}
\end{eqnarray}
\noindent which is a quadratic function of $\boldsymbol{\phi}$.

Next, we need to transform the radar SNR, $\gamma_R$, in (\ref{eqn:radar_snr2}) to make $({\mathbb P}_2)$ solvable. Recall that the radar channel is $\mathbf C_T = \beta \mathbf H_{ul} \mathbf \Phi \mathbf a_I(\psi_{a},\psi_{e}) \mathbf a_I^{T}(\psi_{a},\psi_{e}) \mathbf \Phi \mathbf  H_{dl}$. By denoting $\mathbf U = \mathbf \Phi \mathbf a_I(\psi_{a},\psi_{e}) \mathbf a_I^{T}(\psi_{a},\psi_{e}) \mathbf \Phi$, $\gamma_R$ can be re-arranged as
\begin{eqnarray}
	&&\gamma_R= \tr{ \mathbf C_{T}^H \mathbf C_{T} \left[ \mathbf w \mathbf w^H + \mathbf W_n \mathbf W_n^H \right]  }/ \sigma_R^2 \nonumber \\&&= \beta^2/\sigma_R^2 \tr{  \mathbf U^H \mathbf  H_{ul}^H \mathbf  H_{ul} \mathbf  U \mathbf  H_{dl} (\mathbf w \mathbf w^H + \mathbf W_n \mathbf W_n^H) \mathbf  H_{dl}^H} \nonumber \\&&\overset{(a)}{=} \beta^2/\sigma_R^2 \mathbf u^H \mathbf Z \mathbf u, \label{eqn:gamma_R_new}
\end{eqnarray}
\noindent where in step (a), $\tr{  \mathbf U^H \mathbf A \mathbf U \mathbf B^T} = \mathbf u^H (\mathbf B \otimes \mathbf A) \mathbf u$ is referred to, $\mathbf Z = [\mathbf  H_{dl} (\mathbf w \mathbf w^H + \mathbf W_n \mathbf W_n^H) \mathbf  H_{dl}^H]^T \otimes (\mathbf  H_{ul}^H \mathbf  H_{ul})$, and $\mathbf u = \text{vec}(\mathbf U)$. Note that both $\mathbf U$ and $\mathbf u$ are quadratic in $\mathbf \Phi$ or $\boldsymbol{\phi}$. To create a lower bound for $\gamma_R$ that is quadratic in $\boldsymbol{\phi}$, we need to have a bound which is linear in $\mathbf u$. For this we invoke MM \cite{Sun2017majorization} as
\begin{eqnarray}
	&&\gamma_R = \beta^2/\sigma_R^2 \mathbf u^H \mathbf Z \mathbf u \nonumber \\ &&\geq \tilde \gamma_R = \beta^2/\sigma_R^2 (\mathbf u^H \mathbf Z \mathbf u_t + \mathbf u_t^H \mathbf Z \mathbf u - \mathbf u_t^H \mathbf Z \mathbf u_t), \label{eqn:tilde_gamma_R}
\end{eqnarray}
\noindent where $\mathbf u_t = \text{vec}[\mathbf \Phi_t \mathbf a_I(\psi_{a},\psi_{e}) \mathbf a_I^{T}(\psi_{a},\psi_{e}) \mathbf \Phi_t]$, where $\mathbf \Phi_t$ is the solution of $\mathbf \Phi$ in $t$-th/previous iteration,  the current iteration index is $(t+1)$, and $\tilde \gamma_R$ is the surrogate function for $\gamma_R$. In addition, the first term of $\tilde \gamma_R$ is re-expressed as
\begin{eqnarray}
	&&\mathbf u^H \mathbf Z \mathbf u_t \overset{(b)}{=} \text{vec}(\mathbf U)^H \text{vec}(\mathbf V) = \tr{\mathbf U^H \mathbf V} \nonumber \\&& \overset{(c)}{=} \tr{\mathbf \Phi^H \mathbf a_I^{\ast}(\psi_{a},\psi_{e}) \mathbf a_I^{H}(\psi_{a},\psi_{e}) \mathbf \Phi^H \mathbf V} \nonumber \\&&= \boldsymbol \phi^H \{[\mathbf a_I^{\ast}(\psi_{a},\psi_{e}) \mathbf a_I^{H}(\psi_{a},\psi_{e})] \circ \mathbf V^T\} \boldsymbol \phi^{\ast},
\end{eqnarray}
\noindent where  $\mathbf u = \text{vec}(\mathbf U)$; $\mathbf V$ is set such that $\text{vec}(\mathbf V) = \mathbf Z \mathbf u_t$;  $\mathbf U = \mathbf \Phi \mathbf a_I(\psi_{a},\psi_{e}) \mathbf a_I^{T}(\psi_{a},\psi_{e}) \mathbf \Phi$. Similarly, the second term in $\tilde \gamma_R$ can be re-written as
\begin{eqnarray}
	\mathbf u_t^H \mathbf Z \mathbf u = \boldsymbol \phi^T \{[\mathbf a_I(\psi_{a},\psi_{e}) \mathbf a_I^{T}(\psi_{a},\psi_{e})] \circ \mathbf Y^T\} \boldsymbol \phi,
\end{eqnarray}
\nonumber where the matrix $\mathbf Y$ satisfies $\text{vec}(\mathbf Y) = \mathbf Z^T \mathbf u_t^{\ast}$. Thereby, $\tilde \gamma_R$ in (\ref{eqn:tilde_gamma_R}) can be written as
\begin{eqnarray}
	\tilde \gamma_R = \boldsymbol \phi^H \mathbf L_2 \boldsymbol \phi^{\ast} + \boldsymbol \phi^T \mathbf L_3 \boldsymbol \phi -\beta^2/\sigma_R^2   \mathbf u_t^H \mathbf Z \mathbf u_t, \label{eqn:tilde_gamma_R_new}
\end{eqnarray}
\noindent where the third term is not relevant to the variable $\boldsymbol \phi$, and
\begin{eqnarray}
	&&\mathbf L_2 = \beta^2/\sigma_R^2 \{[\mathbf a_I^{\ast}(\psi_{a},\psi_{e}) \mathbf a_I^{H}(\psi_{a},\psi_{e})] \circ \mathbf V^T\},  \\ &&\mathbf L_3 = \beta^2/\sigma_R^2 \{[\mathbf a_I(\psi_{a},\psi_{e}) \mathbf a_I^{T}(\psi_{a},\psi_{e})] \circ \mathbf Y^T\}.
\end{eqnarray}
 
So far, the objective (\ref{eqn:obj2}) and constraint (\ref{eqn:radar_snr2}) in $	({\mathbb P}_2)$ are transformed into non-fractional quadratic form. By invoking (\ref{eqn:obj2_new}) and (\ref{eqn:tilde_gamma_R_new}), $	({\mathbb P}_2)$ can be re-written as

\begin{subequations} \label{eqn:problem2_newer}
	\begin{eqnarray} 
	\!\!\!\!\!\!\!\!\!\!	(\overline{\mathbb P}_2) \quad	\max_{ \boldsymbol{\phi}}&&	\!\!\!\!\!\boldsymbol \phi^T (\mathbf L_1 +\overline{\mathbf L}_1)\boldsymbol \phi^{\ast} +\Re{[\boldsymbol{\phi}^T (\boldsymbol \eta-\boldsymbol \mu - \overline{\boldsymbol \mu})]}	\label{eqn:obj2_newer}\\ \mathrm{s.t.} && \!\!\!\!\! 
		|\boldsymbol \phi_{n,1}| = 1,\;\; \forall n = 1, \cdots, N\label{eqn:unit_modu2_newer} \\ && \!\!\!\!\! \boldsymbol \phi^H \mathbf L_2 \boldsymbol \phi^{\ast} + \boldsymbol \phi^T \mathbf L_3 \boldsymbol \phi\geq  \gamma_{R,th}' \label{eqn:radar_snr2_newer}
	\end{eqnarray}
\end{subequations}
\noindent  where $\gamma_{R,th}' = \gamma_{R,th} + \beta^2/\sigma_R^2   \mathbf u_t^H \mathbf Z \mathbf u_t $. Now, $(\overline{\mathbb P}_2)$ is a quadratic programming problem with UMC on the elements of the variable $\boldsymbol \phi$ as (\ref{eqn:unit_modu2_newer}).
{$(\overline{\mathbb P}_2)$ can be solved via the SDR method by rewriting it as follows \cite{Li2023anIRS}:
\begin{subequations} \label{eqn:problem2_newest}
	\begin{eqnarray} 
		\!\!\!\!\!\!\!\!\!\!	(\overline{\overline{\mathbb P}}_2) 	\max_{\mathbf R_1, \mathbf R_2, \boldsymbol{\phi}}&&	\!\!\! \!\!\! \tr{( \mathbf L_1 +\overline{\mathbf L}_1)\mathbf R_1^{\ast}} +\Re{[\boldsymbol{\phi}^T (\boldsymbol \eta-\boldsymbol \mu - \overline{\boldsymbol \mu})]}	\label{eqn:obj2_newest}\\ \mathrm{s.t.} && \!\!\!\!\!\!
		|[\mathbf R_1]_{n,n}| \leq 1, \forall n = 1, \cdots, N \label{eqn:R1} \\&&  \!\!\!\!\!\!
		|[\mathbf R_2]_{n,n}| \leq 1, \forall n = 1, \cdots, N \label{eqn:R2} \\ &&  \!\!\!\!\!\!\tr{\mathbf L_2 \mathbf R_2^{\ast}}  + \tr{\mathbf L_3 \mathbf R_2}   \geq   \gamma_{R,th}' \label{eqn:radar_snr2_newest} \\ && \!\!\!\!\!\! \mathbf R_1 \succcurlyeq \boldsymbol{\phi} \boldsymbol{\phi}^H \label{eqn:R1_relax}
		\\ && \!\!\!\!\!\! \mathbf R_2 \succcurlyeq \boldsymbol{\phi} \boldsymbol{\phi}^T \label{eqn:R2_relax}
	\end{eqnarray}
\end{subequations} 
where $\mathbf R_1 = \boldsymbol{\phi} \boldsymbol{\phi}^H$ and $\mathbf R_2 = \boldsymbol{\phi} \boldsymbol{\phi}^T$.
\noindent The latter two assignments represent constraints on  $\mathbf R_1$ and $\mathbf R_2$ which  
 are not allowed in the CVX toolbox \cite{cvx}.  To make the problem (\ref{eqn:problem2_newest}) solvable in CVX, these two constraints are  relaxed as matrix inequalities in (\ref{eqn:R1_relax}) and (\ref{eqn:R2_relax}). For example,  (\ref{eqn:R1_relax}) means that $\mathbf R_1 - \boldsymbol{\phi} \boldsymbol{\phi}^H$ is a positive semidefinite matrix. 
 To improve the performance, we can use Gaussian randomization to acquire the solution of $\boldsymbol{\phi}$. That is, we can generate multiple random vectors 
{whose Gram is $\mathbf R_1$} 
 then element-wise normalize them, and  choose as the solution the vector that  maximizes the objective (\ref{eqn:obj2_newest}) while satisfying the constraint (\ref{eqn:radar_snr2_newest}).
 \noindent The obtained  $\boldsymbol{\phi}$ is substituted into sub-problem 1 of the next iteration. The overall optimization algorithm for jointly designing $\mathbf w$, $\mathbf W_n$ and  $\mathbf \Phi$, is summarized as Algorithm \ref{alg:alt_opt}, where $\mathrm{obj}^{(t+1)}$ is the objective, or secrecy rate $(R_u-R_{te})$, obtained in the $(t+1)$-th iteration, and $\varepsilon$ is the indicator of error tolerance. The quadratic transform \cite{Shen2018fractional} based IRS parameter update method presented in this section is referred to as ``qtSDR" for short. In addition, to solve the design problem (\ref{eqn:orig_prob}), we invoke  qtSDR along with the waveform and AN design method proposed in Section \ref{sec:waveform_AN_design}, as summarized in Algorithm 1.} 
 

\begin{algorithm}\label{alg:alt_opt}
	\SetAlgoLined
	\KwResult{Return $\mathbf w$, $\mathbf W_n$ and  $\mathbf \Phi$.}
	\textbf{Initialization:} $\mathbf \Phi = \mathbf \Phi_0$, $\mathbf w = \mathbf w_0$, $\mathbf W_n = \mathbf W_{n,0}$, $\gamma_u =\gamma_{u,0}$, $\alpha_u =\alpha_{u,0}$, $\gamma_{te} =\gamma_{te,0}$, $\alpha_{te} =\alpha_{te,0}$,  $t=0$\;
	\While{ $\!\!\!(1)$ }{
		
		\textbf{ \emph{// Auxiliary variables update}}
		
		$\alpha_u = {\sqrt{1\!+\!\gamma_u}| \mathbf c_u^T \mathbf w| }/{\left( | \mathbf c_u^T \mathbf w|^2 \!+\! || \mathbf c_u^T \mathbf W_n||^2 \!+\! \sigma_u^2 \right)}$\;
		
		$\gamma_u = {| \mathbf c_u^T \mathbf w|^2 }/{\left( || \mathbf c_u^T \mathbf W_n||^2 + \sigma_u^2 \right)}$\;
		
		$\alpha_{te} = {\sqrt{1 \!+\!\gamma_{te}}| \mathbf c_{te}^T \mathbf w| }/{\left( | \mathbf c_{te}^T \mathbf w|^2 \!\!+\! || \mathbf c_{te}^T \mathbf W_n||^2 \!\!+ \!\sigma_{te}^2 \right)}$\;
		
		$\gamma_{te} = {| \mathbf c_{te}^T \mathbf w|^2 }/{\left( || \mathbf c_{te}^T \mathbf W_n||^2 + \sigma_{te}^2 \right)}$\;

		\textbf{ \emph{// First sub-problem} }
		
		Solve $(\overline{\mathbb P}_1)$ of (\ref{eqn:problem1_new}) to obtain $\mathbf w$ and $\mathbf R_{W_n}$\;

  Obtain $\mathbf W_n$ as the 
		square root matrix of $\mathbf R_{W_n}$ \;

		\textbf{  \emph{//  Second sub-problem {via qtSDR}} }
		
		Substitute $\mathbf w$, $\mathbf W_n$, found  in the first sub-problem  into the second sub-problem\;
		
		Obtain $\boldsymbol{\phi}$ by solving $(\overline{\overline{\mathbb P}}_2)$ in (\ref{eqn:problem2_newest})\;
		
		$\mathbf \Phi = \text{diag}(\boldsymbol\phi)$\;

		\textbf{ \emph{// Condition of termination}}
		
		\If{$\left((t\geq t_{\max})\;||\;(\big|\!\left[ \mathrm{obj}^{(t+1)} \!\!-\! \mathrm{obj}^{(t)}\right]\!/\mathrm{obj}^{(t)}\!\big| \leq \varepsilon)\right)$}{
			\texttt{Break}\;
		}
		
		$t = t+1$;
		
	}
	\caption{Secrecy rate maximization {via qtSDR.}}
\end{algorithm}
\vspace{-0mm}

\section{Closed-Form Expression Based Low-Complexity IRS Parameter Update Scheme} \label{sec:closed_form}

The qtSDR approach, presented in the previous section  
relaxes the UMC constraint, which may lead to loss in the secrecy rate. 
     Further, for solving  $(\overline{\overline{\mathbb P}}_2)$ of (\ref{eqn:problem2_newest}) involves IPM for updating the IRS parameter $\boldsymbol{\phi}$.
       IPM has high complexity when the size of $\boldsymbol{\phi}$ ($N$) is large, which  hinders the scalability of qtSDR. To bypass the aforementioned drawbacks, in this section we propose a closed-form expression based method for updating $\boldsymbol{\phi}$ in each iteration, that takes into account the UMC and has low-complexity.

       {Inside each iteration of the algorithm, at first we solve for the waveform and AN as the first sub-problem block of Algorithm \ref{alg:alt_opt} shows. Then in the second sub-problem block, we use a closed-form expression to update the values of the IRS parameters, and this expression will invoke the value of the solved IRS parameters in previous outermost iteration. We denote the index of the outermost iteration as $t$.}

       MM methods enable new ways to convexify non-convex problems \cite{Sun2017majorization,He2022qcqp}, i.e.  by creating tight convex  lower{/upper} bounds of the original non-convex functions, and then iteratively  maximizing{/minimizing} those bounds. In our problem, our aim is to maximize the secrecy rate {(see (\ref{eqn:obj2_newer}))}, while keeping the SNR at the radar receiver above a threshold {(see (\ref{eqn:radar_snr2_newer}))}. Both (\ref{eqn:obj2_newer}) and (\ref{eqn:radar_snr2_newer}) contain quadratic functions of $\boldsymbol{\phi}$. In addition, the  UMC on the elements of IRS parameter in (\ref{eqn:unit_modu2_newer}) imposes more non-convexity to the problem of (\ref{eqn:problem2_newer}).

    Firstly, by invoking MM \cite{Sun2017majorization}, we create lower bounds for the quadratic functions in (\ref{eqn:problem2_newer}), which are linear in $\boldsymbol{\phi}$, i.e., 
    \addtocounter{equation}{+1}
\begin{eqnarray}
    \boldsymbol \phi^T (\mathbf L_1 +\overline{\mathbf L}_1)\boldsymbol \phi^{\ast} \!\!\!\!\!\!\!\!\!\!&&\geq \boldsymbol \phi^T_t (\mathbf L_1 +\overline{\mathbf L}_1)\boldsymbol \phi^{\ast} + \boldsymbol \phi^T (\mathbf L_1 +\overline{\mathbf L}_1)\boldsymbol \phi^{\ast}_t \nonumber \\ \!\!\!\!\!\!\!\!\!\!&& - \boldsymbol \phi^T_t (\mathbf L_1 +\overline{\mathbf L}_1)\boldsymbol \phi_t^{\ast},
\end{eqnarray}
\begin{eqnarray}
    \boldsymbol \phi^H \mathbf L_2 \boldsymbol \phi^{\ast} \geq 2 \boldsymbol \phi^H_t \mathbf L_2 \boldsymbol \phi^{\ast} - \boldsymbol \phi^H_t \mathbf L_2 \boldsymbol \phi^{\ast}_t,
\end{eqnarray}
\begin{eqnarray}
    \boldsymbol \phi^T \mathbf L_3 \boldsymbol \phi \geq 2 \boldsymbol \phi^T_t \mathbf L_3 \boldsymbol \phi - \boldsymbol \phi^T_t \mathbf L_3 \boldsymbol \phi_t,
\end{eqnarray}
\noindent where the current iteration index is $t+1$, and $\boldsymbol \phi_t$ is the {obtained value for}
$\boldsymbol{\phi}$ in {the previous/$t$-th outermost iteration.} 

Thereby, the  problem $(\overline{\mathbb P}_2)$ in (\ref{eqn:problem2_newer}) is re-reformulated as

\addtocounter{equation}{+1}
    \begin{eqnarray}   \label{eqn:P2_tilde}
    \!\!\!\!\!\!\!\!\!\!(\tilde {\mathbb P}_2) \quad	\max_{ \boldsymbol{\phi}}&&	\!\!\!\!\!\boldsymbol \phi^T_t (\mathbf L_1 +\overline{\mathbf L}_1)\boldsymbol \phi^{\ast} + \boldsymbol \phi^T (\mathbf L_1 +\overline{\mathbf L}_1)\boldsymbol \phi^{\ast}_t  \nonumber \\ \!\!\!\!\!\!\!\!\!\!\!\!\!\!\!\!\!\!&&\!\!\!\!\!
    - \boldsymbol \phi^T_t (\mathbf L_1 +\overline{\mathbf L}_1)\boldsymbol \phi_t^{\ast} +\Re{[\boldsymbol{\phi}^T (\boldsymbol \eta-\boldsymbol \mu - \overline{\boldsymbol \mu})]}	\label{eqn:obj2_mm} 
    \end{eqnarray}
    \vspace{-7mm}
    \begin{eqnarray}
    \!\!\!\!  \!\!\!\!\!\!\!\!\! \!\!\!\!\!\!\!\!\!\!\!\!\!\!\!\!\!\!\!\!\!\!\!\!\!\!\!\mathrm{s.t.} \!\!&& \!\!\!\!\! 
		|\boldsymbol \phi_{n,1}| = 1,\;\; \forall n = 1, \cdots, N\label{eqn:unit_modu2_mm} 
  \end{eqnarray}
  \vspace{-10mm}

  \addtocounter{equation}{+1}
  \begin{eqnarray}
  \!\!\!\!\!\!\!\!\!\!\!\!\!\!\!\!\!\!&& \!\!\!\!\! 2 \boldsymbol \phi^H_t \mathbf L_2 \boldsymbol \phi^{\ast} - \boldsymbol \phi^H_t \mathbf L_2 \boldsymbol \phi^{\ast}_t + 2\boldsymbol \phi^T_t \mathbf L_3 \boldsymbol \phi  \nonumber\\ \!\!\!\!\!\!\!\!\!\!\!\!\!\!\!\!\!\!&& \!\!\!\!\! -\boldsymbol \phi^T_t \mathbf L_3 \boldsymbol \phi_t \geq  \gamma_{R,th}' \label{eqn:radar_snr2_mm}
    \end{eqnarray}

    For notation simplicity, we denote the objective in (\ref{eqn:obj2_mm}) as $g_0(\boldsymbol{\phi})$, and

\vspace*{-2mm}
    \begin{small}   
    \begin{eqnarray}
        g_1(\boldsymbol{\phi}) = 2 \boldsymbol \phi^H_t \mathbf L_2 \boldsymbol \phi^{\ast} \!-\! \boldsymbol \phi^H_t \mathbf L_2 \boldsymbol \phi^{\ast}_t \!+\! 2\boldsymbol \phi^T_t \mathbf L_3 \boldsymbol \phi   \!-\!\boldsymbol \phi^T_t \mathbf L_3 \boldsymbol \phi_t \!-\! \gamma_{R,th}'.
    \end{eqnarray}
    \end{small}

Thereby, the problem $(\tilde {\mathbb P}_2)$ can be re-written as
\begin{subequations} \label{eqn:problem2_simple}
	\begin{eqnarray} 
	\!\!\!\!\!\!\!\!\!\!	(\tilde {\mathbb P}_2) \quad	\max_{ \boldsymbol{\phi}}&&	\!\!\!\!\! g_0(\boldsymbol{\phi})	\label{eqn:obj2_simple}\\ \mathrm{s.t.} && \!\!\!\!\! 
		|\boldsymbol \phi_{n,1}| = 1,\;\; \forall n = 1, \cdots, N \\ && \!\!\!\!\! g_1(\boldsymbol{\phi}) \geq 0 \label{eqn:radar_snr2_simple}
	\end{eqnarray}
\end{subequations}
So far, we have obtained an LP problem subject to UMC similar to \cite{He2022qcqp}. In \cite{He2022qcqp}, an objective minimization problem with upper bound constraints is studied, while in our paper an objective maximization problem with lower bound constraints is studied. Similar to \cite{He2022qcqp}, we formulate and solve the Lagrangian dual problem of the original LP problem with UMC as follows.
The Lagrangian dual function of the problem in (\ref{eqn:problem2_simple}) can be written as
\begin{eqnarray} \label{eqn:lagrangian}
   \mathcal L(\rho) = \max_{\boldsymbol{\phi} \in \Omega } \quad \Re{[g_0(\boldsymbol{\phi}) + \rho g_1(\boldsymbol{\phi})]}
\end{eqnarray}
\noindent where $\Omega = \{ \boldsymbol{\phi} \in \mathbb C^N \big| \;\; |\boldsymbol \phi_{n,1}| = 1,\;\; \forall n = 1, \cdots, N\}$, $\rho$ is the Lagrangian multiplier and non-negative. The Lagrangian dual problem of $(\tilde {\mathbb P}_2)$ is 
\begin{eqnarray}
    L^{\star} = \min_{\rho \geq 0} \mathcal L(\rho)
\end{eqnarray}
The Lagrangian $\mathcal L(\rho)$ in (\ref{eqn:lagrangian}) is linear in $\boldsymbol{\phi}$ noting that $g_0(\boldsymbol{\phi})$ and $g_1(\boldsymbol{\phi})$ are both linear functions of $\boldsymbol{\phi}$, therefore the optimal $\mathcal L(\rho)$ can be obtained by letting
\begin{eqnarray} \label{eqn:phi_update}
    \boldsymbol{\phi}(\rho) = \exp{[j \arg(\boldsymbol \nu(\rho) + \boldsymbol \kappa^{\ast}(\rho))]},
\end{eqnarray}

\noindent where 
\begin{eqnarray}
   \!\!\!\!\!\!\! \boldsymbol{\nu}(\rho) \!\!\!\!\!&=&\!\! \!\!\!\left(\boldsymbol \phi^T_t (\mathbf L_1 +\overline{\mathbf L}_1) + 2 \rho \boldsymbol \phi_t^H \mathbf L_2\right)^T, \label{eqn:nu} \\\!\!\!\!\!\!\!
    \boldsymbol{\kappa}(\rho) \!\!\!\!\!&=& \!\!\!\!\!\left[\boldsymbol \phi^H_t (\mathbf L_1 +\overline{\mathbf L}_1)^T + 2 \rho \boldsymbol \phi_t^T \mathbf L_3 + (\boldsymbol \eta-\boldsymbol \mu - \overline{\boldsymbol \mu})^T \right]^T\!\!\!.\label{eqn:kappa}
\end{eqnarray}
The derivation of Eq. (\ref{eqn:phi_update}) can be found in Appendix (\ref{sec:phi_update_qtMM}).
Notice that the IRS parameter vector $\boldsymbol{\phi}$ is now a function of $\rho$. A properly chosen $\rho$ can ensure that  the objective $g_0(\boldsymbol{\phi}(\rho))$ is maximized while the constraint $g_1(\boldsymbol{\phi}(\rho)) \geq 0$ is satisfied.

{The parameter}
$\rho$ quantifies the importance of the radar SNR constraint. In each iteration, initially $\rho=0$, {indicating  maximization of} the objective  $g_0(\boldsymbol{\phi}(0))$ with respect to $\boldsymbol{\phi}$ without considering the radar SNR constraint. If {the} obtained solution, say $\boldsymbol{\phi}^{\star}(0)$,   satisfies $g_1(\boldsymbol{\phi}^{\star}(0)) \geq 0$, then $\rho$ is set {to} $0$ in {the} current iteration. In this case, the $\boldsymbol{\phi}^{\star}(0)$  is used as the solution of {the} IRS parameter in {the} current iteration. 
If $g_1(\boldsymbol{\phi}^{\star}(0)) <0$, this means that the radar SNR constraint is not satisfied with the obtained solution and a  positive $\rho$ is necessary.  If $\rho$ is too large, the radar SNR constraint is given too much importance, which might consume too many DoFs of the system, 
thus making it impossible to meet the secrecy rate objective.
In this case, we find an upper bound of $\rho$ at first, say $\rho_{\text{ub}} = 2^x$, where $x$ is the smallest integer that satisfies $g_1(\boldsymbol{\phi}(\rho_{\text{ub}})) = g_1(\boldsymbol{\phi}(2^x)) \geq 0$. Then, a smallest $\rho^{\star}$ is found in the range of $(0, \rho_{\text{ub}})$ using {the} bisection method conditioned on $g_1(\boldsymbol{\phi}(\rho^{\star})) \geq 0$. In this case, the least importance is allocated to the radar SNR constraint while this constraint is met, so that we have reserved most system DoFs to be used for maximizing the objective of secrecy rate. The closed-form expression based IRS parameter update   is summarized as Algorithm \ref{alg:IRS_update}. Algorithm \ref{alg:IRS_update} is called by Algorithm \ref{alg:alt_opt} when the IRS parameter needs to be updated in the second sub-problem {block} in each iteration. The closed-form expression based IRS parameter update method proposed in this section, which invokes quadratic transform \cite{Shen2018fractional},  is named as ``qtMM" for short. 


\begin{algorithm}\label{alg:IRS_update}
	\SetAlgoLined
	\KwResult{Return $\boldsymbol{\phi}^{\star}(\rho^{\star})$ as the IRS parameter.}

		\If{$g_1(\boldsymbol{\phi}^{\star}(0)) \geq 0$}{
			$\rho^{\star} = 0$\;

   \Else{
        $x=x_{\text{ini}}$\;

        \While{
        $g_1(\boldsymbol{\phi}(2^x))) < 0$
        }
        {$x=x+1$\;}

        $\rho_{\text{ub}} = 2^x$\; 
        
        $\rho_{\text{lb}} = 2^{x-1}$\;
        $\rho_{\text{mid}} =  ({\rho_{\text{lb}} + \rho_{\text{ub}}})/{2} $\;
        
        \While{ \big[ \!\! $g_1(\boldsymbol{\phi}(\rho_{\text{mid}}))) < 0$ \! \!\!$||$ \!$g_1(\boldsymbol{\phi}(\rho_{\text{mid}}))) > \epsilon_{\text{in}}$ \!\!\big]}
        {
        
        \If{$g_1(\boldsymbol{\phi}(\rho_{\text{mid}}))) < 0$}
        {
        $\rho_{\text{lb}} = \rho_{\text{mid}}$\;

        \Else{ $\rho_{\text{ub}} = \rho_{\text{mid}}$\; }
        }
        
        $\rho_{\text{mid}} =  ({\rho_{\text{lb}} + \rho_{\text{ub}}})/{2} $\;
        
        }
        
    $\rho^{\star} = \rho_{\text{mid}}$\;
  
  }
		}

  $\boldsymbol{\phi}^{\star}(\rho^{\star}) = \exp{[j \arg(\boldsymbol \nu(\rho^{\star}) + \boldsymbol \kappa^{\ast}(\rho^{\star}))]}$\;
  
	\caption{{qtMM -} Update of IRS parameter   in $(t+1)$-th/current iteration based on proposed closed-form solution based low-complexity scheme in Section \ref{sec:closed_form}.}
\end{algorithm}
\vspace{-0mm}

\begin{table}[]
	\begin{center}			
		\caption{System Configurations}
		\label{tab:sys_para}
		\begin{tabular}{p{6.2cm}|p{1.8cm}} 
			\hline
			\textbf{Parameter} & \textbf{Value}
			\\\hline
			Error tolerance indicator of Algorithm \ref{alg:alt_opt},  $\varepsilon$ [dB] & $- 20$
			\\\hline
			Maximum number of iterations allowed, $t_{\max}$ & $15$			
			\\\hline
			Rician factors of $\mathbf g$, $\mathbf f$, $\mathbf H_{dl}$, and $\mathbf H_{ul}$ [dB] & $20$
			\\\hline
			  Noise power at the radar receiver,  $\sigma_R^2$ [dBm]&$0$
			\\\hline
			Noise power at the communication user, $\sigma_u^2$ [dBm]&$0$
			\\\hline
			Noise power at the target/ED, $\sigma_{te}^2$ [dBm]&$0$
			\\\hline
			  Inter-antenna distance at radar transmitter/receiver, $d$ & $ \lambda/2$
			\\\hline
			Number of  antennas of radar transmitter, $N_T$ & $16$
			\\\hline
			Number of  antennas of radar receiver, $N_R$ & $16$
			\\\hline
			Number of  IRS elements, $N$ & $64$
			\\\hline
			Target coefficient $|\beta|$ [dB] & $-40$
			\\\hline
			SNR threshold at radar receiver $\gamma_{R,th}$ [dB] & $-11$
			\\\hline
		Radar power budget $P_R$ [dBm] & $30$
			\\\hline
	 Elevation and azimuth  angles of  target relative to  IRS $(\psi_e, \psi_a)$ & $(60^{\circ},30^{\circ})$
  \\\hline
   Elevation and azimuth  angles of  user relative to  IRS  & $(-45^{\circ},-45^{\circ})$
  \\\hline
  Azimuth  angle of  user relative to  radar  & $-30^{\circ}$
  \\\hline
  Azimuth  angle of  IRS relative to  radar  & $60^{\circ}$
  \\\hline

			CSI error level  & $0$
     \\\hline
		
		\end{tabular}
	\end{center}
	\vspace{-0mm}
\end{table}

\section{NUMERICAL RESULTS}
In this section, we present numerical results to demonstrate the convergence of the proposed algorithm, {its beamforming performance, its secrecy performance, and its advantage in terms of scalability}. 
%
%

{
While AN helps PLS, it results in  loss of communication signal power. We define  $\omega$ as the ratio of 
user information power  over the total power of user information and AN, and    show, via simulations,  how $\omega$ affects the system performance}.

\begin{figure}[!t]\centering\vspace{-5mm}
	\includegraphics[width=0.42\textwidth]{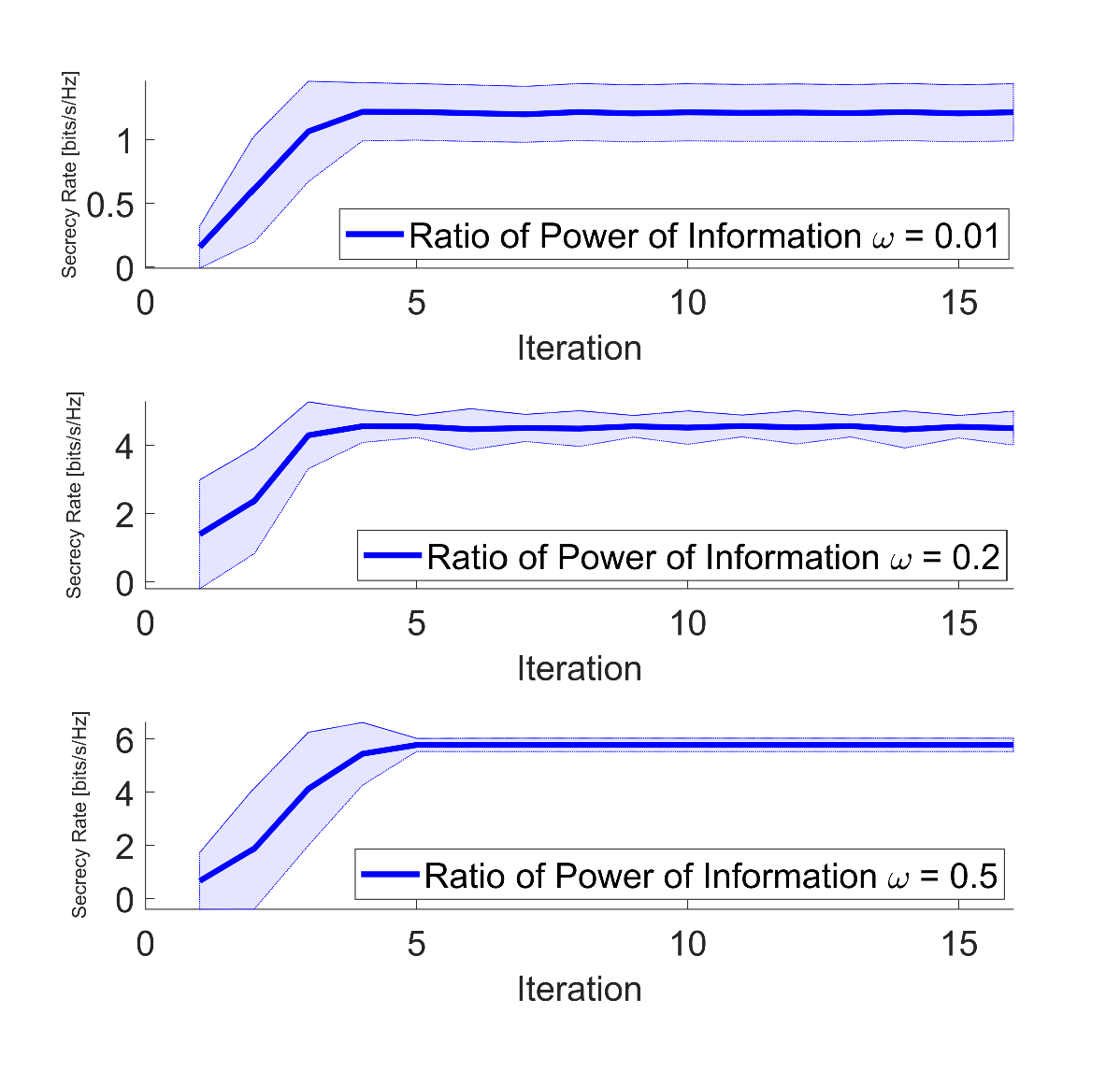}
	\caption{Convergence of the proposed algorithm when the ratio of information power $\omega$ varies.}
	\label{fig:Fig_converge_diff_omega_perfectCSI}\vspace{-3mm}
\end{figure}

The channels are simulated as Rician {fading channels}, i.e., 


\vspace*{-5mm}
\begin{eqnarray}
    \mathbf g = \sqrt{\frac{\kappa_g}{\kappa_g+1}} \mathbf g_{los} + \sqrt{\frac{1}{\kappa_g+1}} \mathbf g_{nlos},
\end{eqnarray}
\noindent where $\mathbf g_{los} \in \mathbb C^{N_T \times 1}$, $\mathbf g_{nlos} \in \mathbb C^{N_T \times 1}$, and $\kappa_g$ are respectively the LoS component, NLoS component and Rician factor of $\mathbf g$. The channels $\mathbf f$, $\mathbf H_{ul}$ and $\mathbf H_{dl}$ are similarly defined.
The parameters of the simulation are taken as shown as in Table \ref{tab:sys_para} {unless otherwise stated}.

\noindent\textit{Convergence - }
Fig. \ref{fig:Fig_converge_diff_omega_perfectCSI} demonstrates the convergence of our proposed  algorithm. The solid blue line represents the mean of the  objective (secrecy rate) versus iteration number, obtained based  $100$ different channel realizations. The light blue shaded area around the solid line shows the variance among the  realizations. From the figure one can see that the objective, $R_u - R_{te}$,  reaches convergence very fast for a wide range of  values of $\omega$.

\begin{figure}[!t]\centering\vspace{-0mm}
	\includegraphics[width=0.38\textwidth]{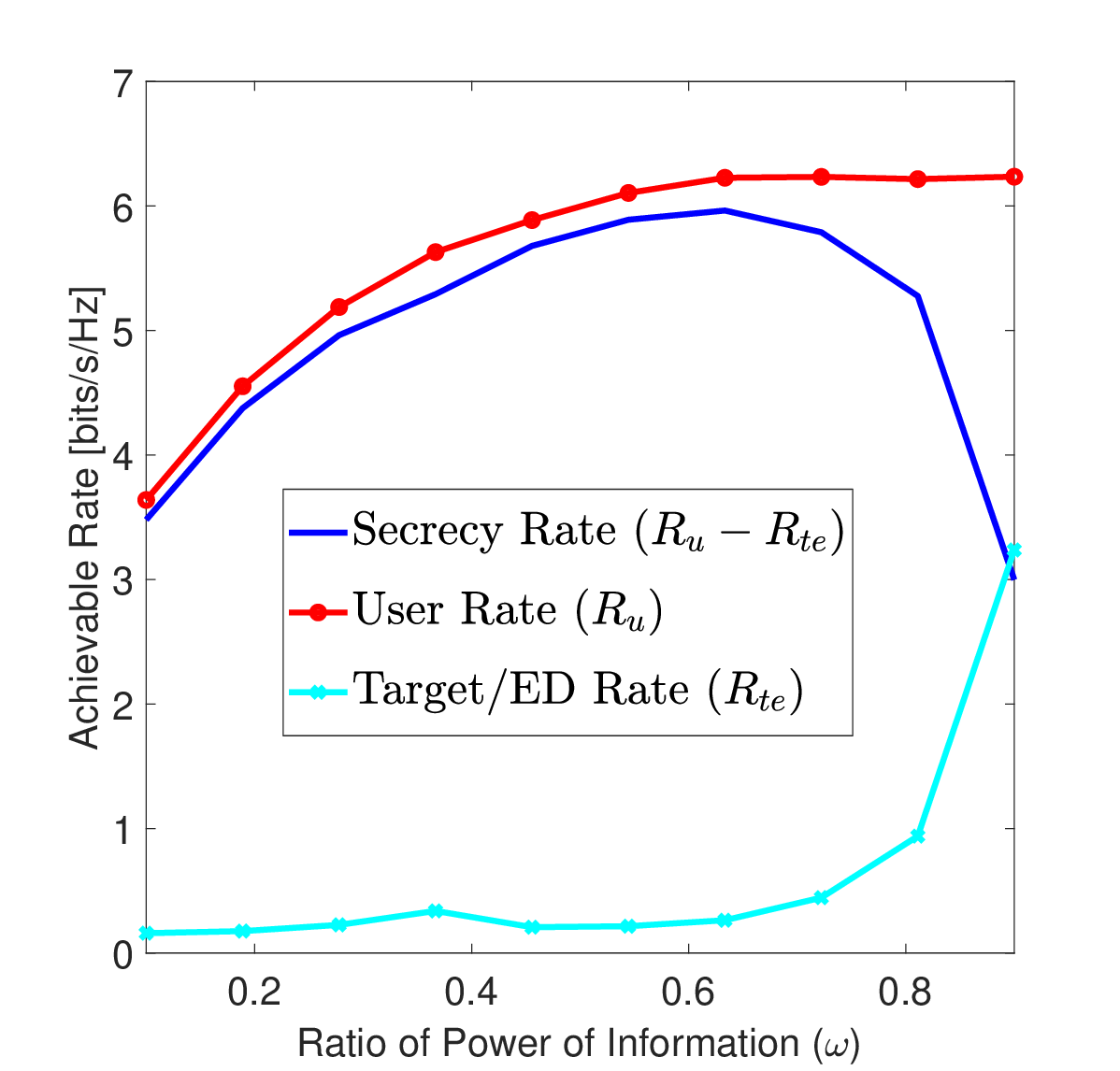}
	\caption{The impact of $\omega$ on the achievable rates.}
	\label{fig:Fig_SR_omega}\vspace{-3mm}
\end{figure}

\noindent\textit{Secrecy performance versus $\omega$} - 
Fig. \ref{fig:Fig_SR_omega} shows the effect of the power ratio of user information, $\omega$, on the achievable rates. When $\omega$ is small, low power is allocated to the user information, and high power to the AN. As a result, the user rate ($R_u$) and rate at the target/ED ($R_{te}$) are both low, and the secrecy rate ($R_u-R_{te}$) is also low. As $\omega$ increases, both $R_u$ and $R_{te}$ increase. $R_u$ increases fast when $\omega$ is small, and  slower when $\omega$ is large. Meanwhile, $R_{te}$ rises slowly when $\omega$ is low, and increases fast when $\omega$ is large. Therefore, with the increase of $\omega$, the secrecy rate, $R_u-R_{te}$, increases at first, since $R_u$ increases faster than $R_{te}$ in the beginning (See Fig. \ref{fig:Fig_SR_omega} when $\omega$ is less than  $0.6$). Afterwards, $R_{te}$ rises faster, so the secrecy rate drops. When $\omega$ is close to $1$, the waveform sent to illuminate the target/ED for the sensing task primarily consists of user information, which leads to  high ED rate, and decreased secrecy rate.

\begin{figure}[!t]\centering\vspace{-0mm}
	\includegraphics[width=0.38\textwidth]{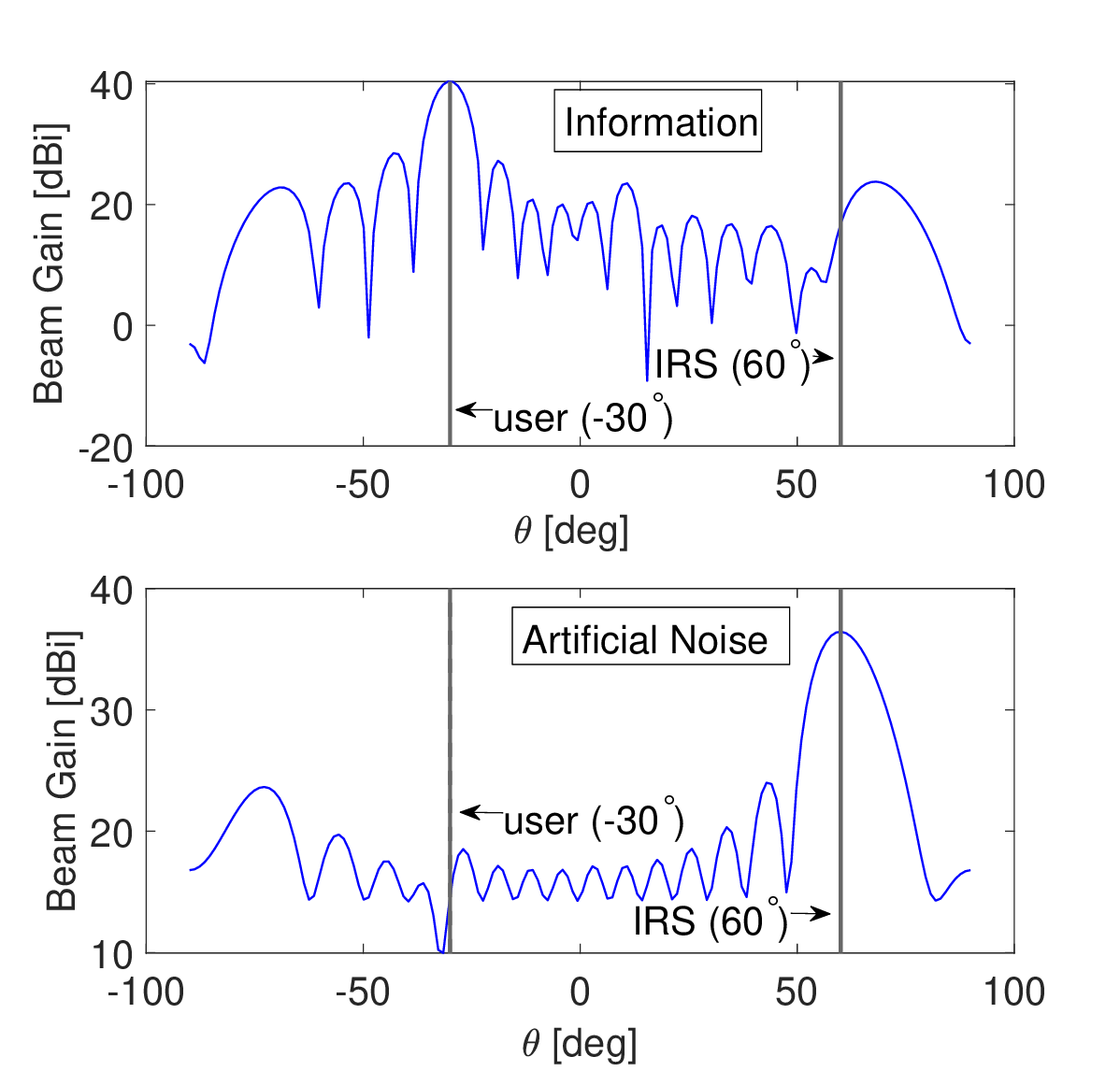}
	\caption{First subplot: radar beampattern contributed by the waveform which contains the information intended for the user; Second subplot: radar beampattern contributed by the AN; $\omega=0.7$, Rician factors: $20$\,dB.
 }
	\label{fig:Fig_radar_beampattern_mm}\vspace{-3mm}
\end{figure}

\begin{figure}[!t]\centering\vspace{-0mm}
	\includegraphics[width=0.38\textwidth]{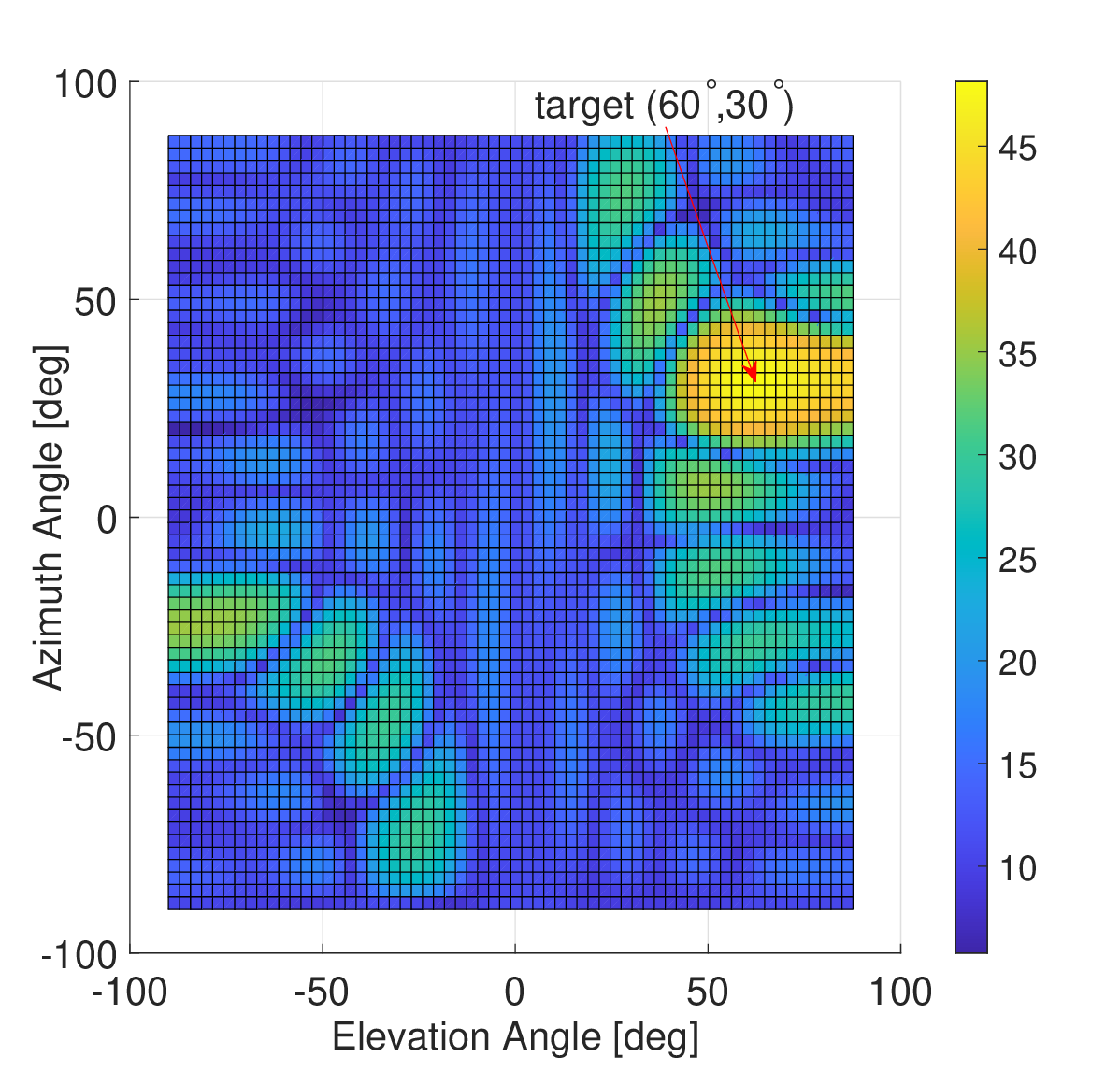}
	\caption{Obtained IRS beampattern; $\omega=0.7$, Rician factors: $20$\,dB.}
	\label{fig:Fig_IRS_beampattern_mm}\vspace{-1mm}
\end{figure}

\noindent 
\noindent\textit{Beampatterns under dominant LoS channel component} - Fig. \ref{fig:Fig_radar_beampattern_mm}    shows the radar beampatterns respectively contributed by the radar waveform that contains the user information and the AN. Fig. \ref{fig:Fig_IRS_beampattern_mm} displays the IRS beampattern. Here, we set $\omega=0.7$ and 
$N = 64$. The Rician factors of channels $\mathbf g$, $\mathbf f$, $\mathbf H_{dl}$ are $20$\,dB, so the LoS components are dominant. It can be  seen from Fig. \ref{fig:Fig_radar_beampattern_mm} that, at the radar, the information bearing signal forms a beam to the direction of the user, and the AN forms one towards the IRS. In addition, the IRS forms a beam to the target, per Fig. \ref{fig:Fig_IRS_beampattern_mm}. Therefore, by our  design, the AN is directed towards the IRS, and upon reflections is directed by the IRS towards the target/ED to simultaneously illuminate and jam it.

\begin{figure}[!t]\centering\vspace{-1mm}
	\includegraphics[width=0.38\textwidth]{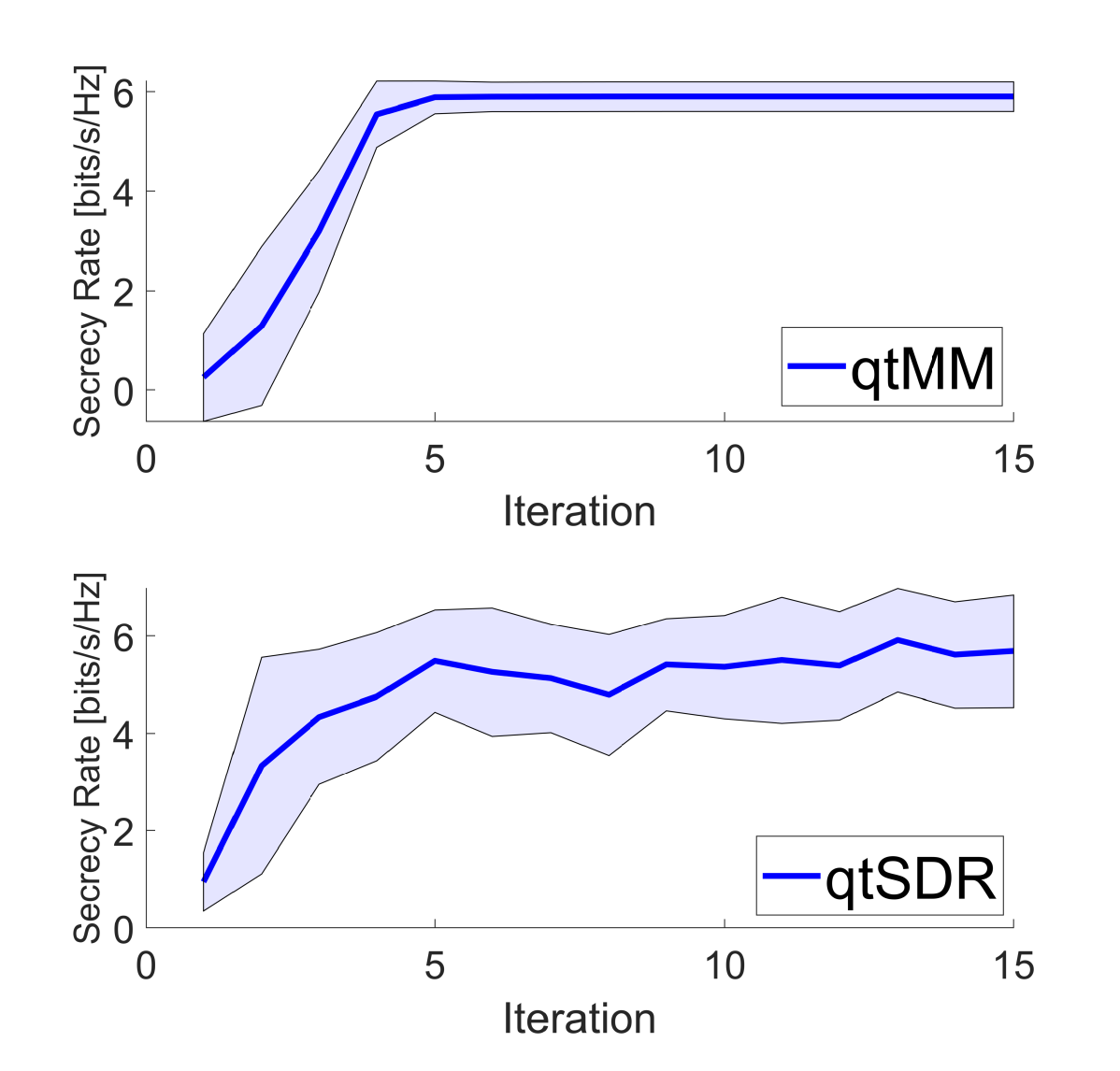}
	\caption{The proposed closed-form expression based IRS parameter update scheme   (qtMM) versus the one in \cite{Li2023anIRS} (qtSDR). }
	\label{fig:Fig_mm_vs_sdr}\vspace{-3mm}
\end{figure}

\noindent 
\noindent\textit{Performance comparison  with qtSDR} - 
Fig. \ref{fig:Fig_mm_vs_sdr} shows the convergence  of the proposed qtMM and that of  qtSDR. Here we set $N=36$ and $\omega=0.5$. It is observed that {the}
 qtMM achieves higher secrecy performance and exhibits less variance during the iterations.
 For both methods, the numbers of iterations to achieve convergence are around $5$. 
 Moreover, for an IRS size $N=36$ on average  qtMM  takes around $6.6$ seconds to achieve convergence, while the qtSDR needs $50$ seconds.

\begin{figure}[!t]\centering\vspace{-0mm}
	\includegraphics[width=0.38\textwidth]{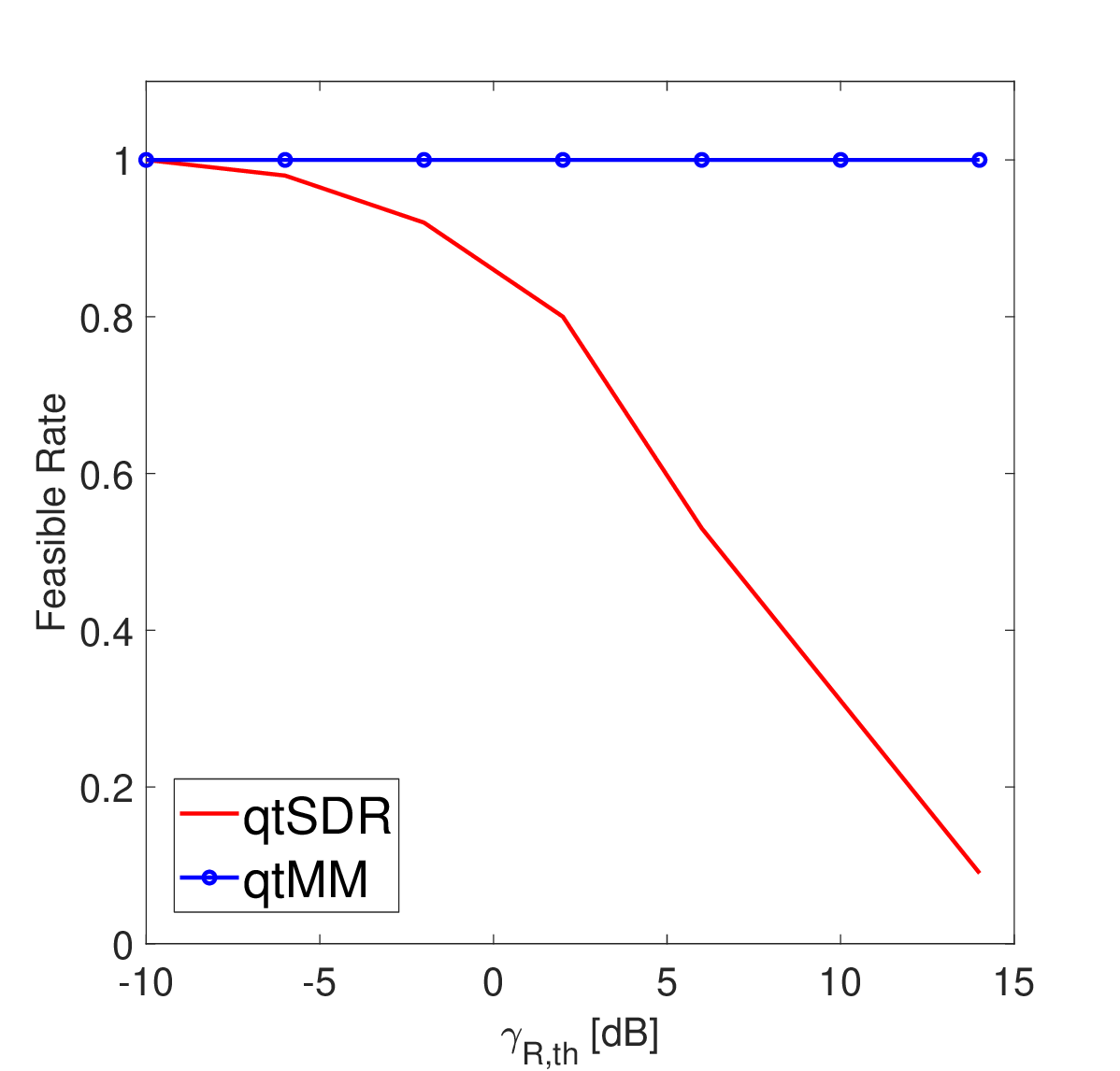}
	\caption{Feasible rates of the proposed methods in the second sub-problem.}
	\label{fig:Fig_feasible_rate_sdr}\vspace{-3mm}
\end{figure}

Let's characterize the feasible rate as the likelihood that the solution obtained for the IRS parameter in the second sub-problem meets the radar SNR constraint, i.e. $\gamma_R(\boldsymbol{\phi})  \geq \gamma_{R,th}$.
Fig. \ref{fig:Fig_feasible_rate_sdr} describes the feasible rates 
of the {solved} IRS parameter   {for $N=25$.}
For the qtSDR, a randomly generated solution, obtained by Gaussian randomization, that statistically attains a  good secrecy rate does not necessarily satisfy the radar SNR constraint in Eq. (\ref{eqn:radar_snr2_newest}). In addition, as the radar SNR threshold $\gamma_{R,th}$ increases, the radar SNR constraint is less likely to be satisfied. For example, when $\gamma_{R,th}=-6$\,dB, the feasible rate of qtSDR, or the probability that the SNR constraint (\ref{eqn:radar_snr2_newest}) is satisfied, is $98$\%. When $\gamma_{R,th}=10$\,dB, the feasible rate of qtSDR decreases to $31$\%.  In addition, the feasible rate of qtMM stays {at} $100$\% as $\gamma_{R,th}$ changes. The explanation is as follows. When $\gamma_{R,th}$ increases, meaning that there is a more strict constraint on the SNR at the radar receiver, the qtMM increases the importance factor ($\rho$) of the radar SNR constraint (\ref{eqn:radar_snr2_simple}) until it is satisfied.

\begin{figure}[!t]\centering\vspace{-0mm}
	\includegraphics[width=0.38\textwidth]{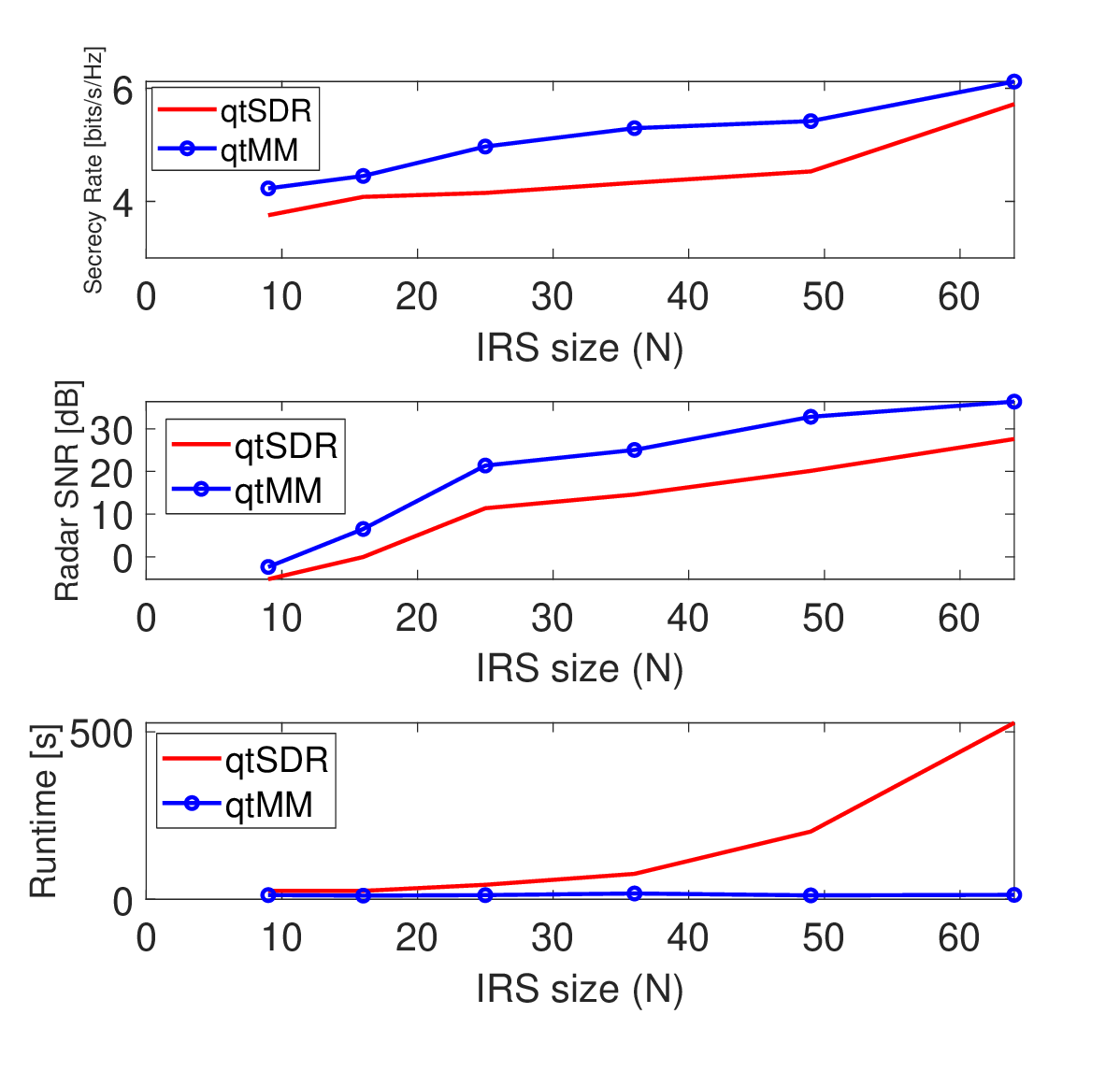}
	\caption{Comparison between proposed qtMM and qtSDR   when the IRS size ($N$) varies.  }
	\label{fig:Fig_sdr_vs_mm_N}\vspace{-3mm}
\end{figure}

Fig. \ref{fig:Fig_sdr_vs_mm_N} compares the qtSDR  and the proposed qtMM in terms of secrecy rate,  radar receiver SNR and average runtime of convergence, as the size of IRS ($N$) changes. From the first subplot, it is observed that the qtMM obtains  better secrecy rate than the qtSDR, and further achieves better radar SNR  per the second subplot.
As shown in the third subplot, while in the low $N$ regime, the average runtimes of the qtMM and the qtSDR are similar, the average runtime of the qtSDR increases fast with the IRS size $(N)$.  The average runtime of   qtMM  stays around $10$\,seconds for $N$ ranging from $9$ to $64$.

\begin{figure}[!t]\centering\vspace{-0mm}
	\includegraphics[width=0.40\textwidth]{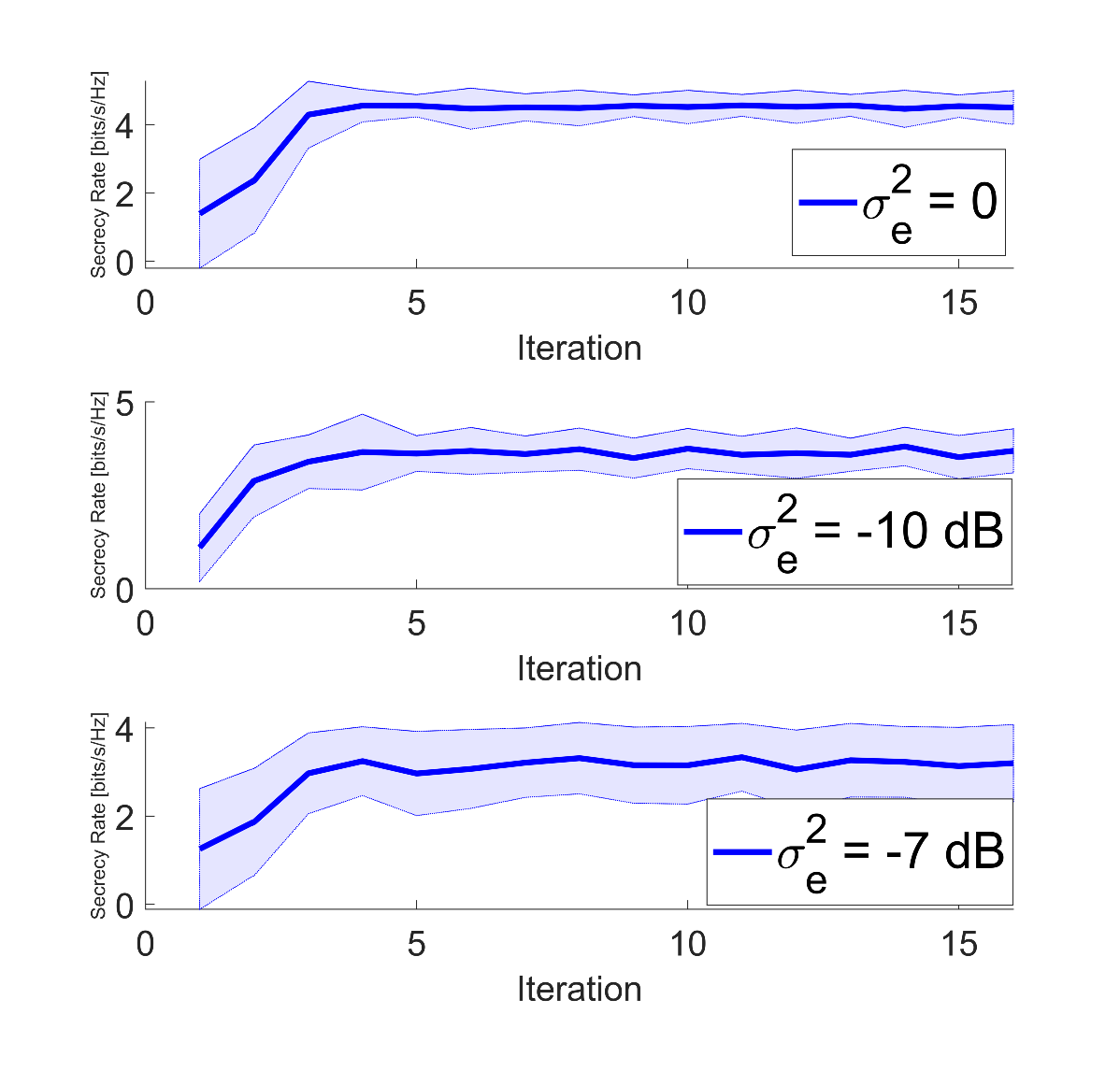}\vspace{-3.5 mm}
	\caption{Convergence of the proposed algorithm for different CSI error levels.}
	\label{fig:Fig_converge_sigma_e2}\vspace{-0mm}
\end{figure}


\noindent\textit{Performance using imperfect channel state information (CSI) - }
The channel estimates will always contain some level of estimation error. On denoting  the $n$-th element of the true channel $\mathbf g$ by $ g_n$, and the   estimated one by ${\hat g}_n$, we can write
\begin{eqnarray}
    g_n =  {\hat g}_n + \epsilon_{g_n},
\end{eqnarray}
\noindent where $\epsilon_{g_n} \sim \mathcal{CN}(0, \frac{\sigma_e^2 \zeta_g}{\kappa_g+1} )$ is the channel error with $\zeta_g$ being the large-scale fading coefficient of the channel $\mathbf g$; ${\hat g}_n \sim \mathcal{CN}(\sqrt{\frac{\kappa_g}{\kappa_g+1}}  g_{los, n}, \frac{1-\sigma_e^2}{1+\kappa_g} \zeta_g)$ where $ g_{los, n}$ is the $n$-th element of $\mathbf g_{los}$. In addition, $\epsilon_{g_n}$ and ${\hat g}_n$ are independent with each other. The imperfect CSI models for $\mathbf f$, $\mathbf H_{ul}$ and $\mathbf H_{dl}$ are defined likewise.

 Fig. \ref{fig:Fig_converge_sigma_e2} shows the convergence of our proposed algorithm as the CSI error level $\sigma_e^2$ varies. In these experiments, $\omega=0.2$, and the Rician factor of the channels are set to $0$\,dB. 
Fig. \ref{fig:Fig_converge_sigma_e2} shows that in all three sub-figures, our proposed method can achieve convergence in a few iterations.  As expected, higher CSI error level results in lower secrecy rate. For example, compared to the $\sigma_e^2 = 0$ case where genie-aided perfect CSI is available, the secrecy rates of $\sigma_e^2 = -10$\,dB case and $\sigma_e^2 = -7$\,dB case are decreased by $19\%$ and $30\%$, respectively. 
When the Rician factors are $0$\,dB,  our proposed qtMM achieves convergence even if $\sigma_e^2$ is as large as $-0.04$\,dB (dimensionless value of $0.99$). The secrecy rate is around $1$\,bits/s/Hz in that case. 

\begin{figure}[!t]\centering\vspace{-0mm}
	\includegraphics[width=0.38\textwidth]{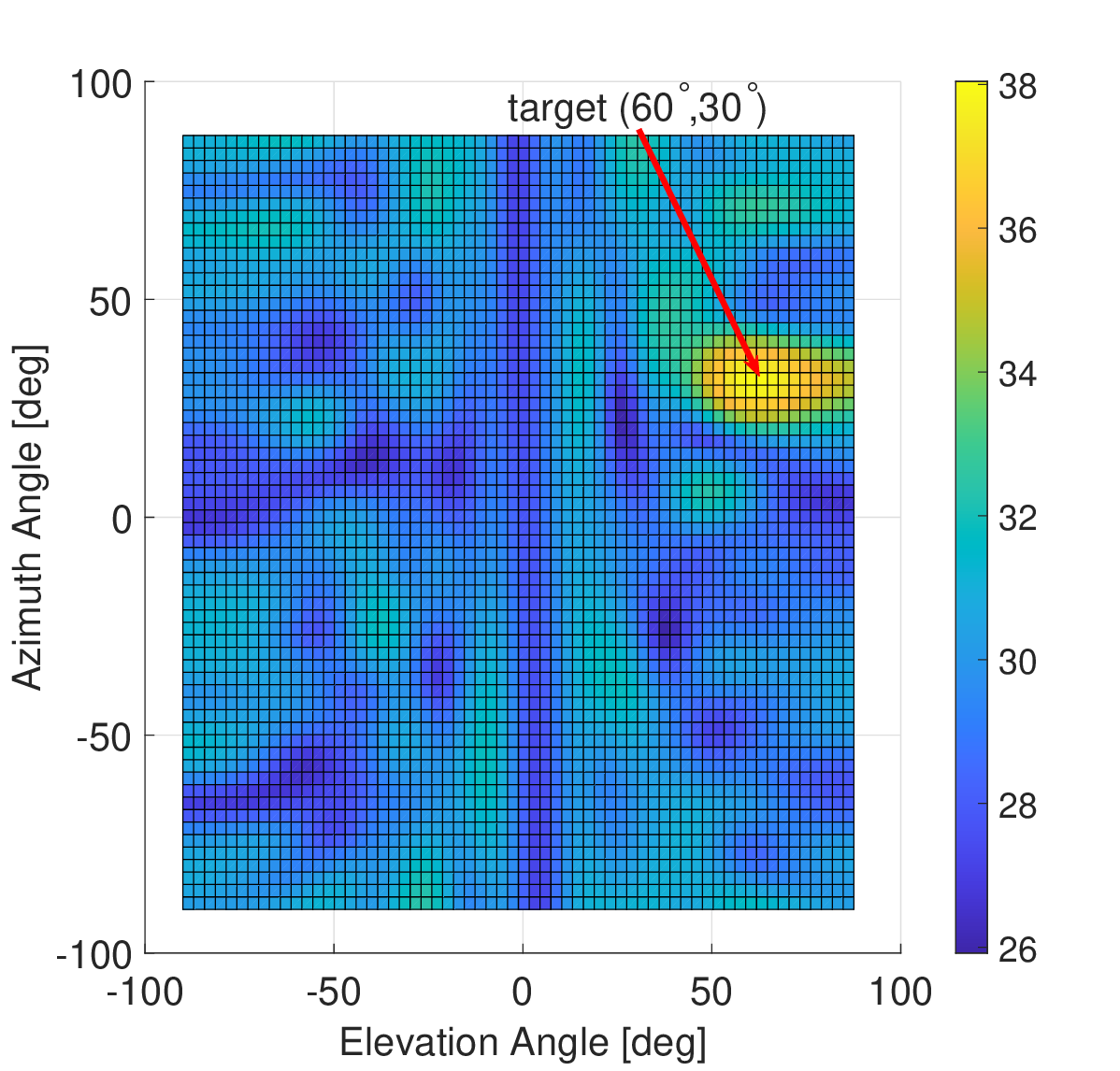}
	\caption{Obtained IRS beampattern; $\omega=0.5$, Rician factors: $-20$\,dB, $\sigma_e^2=0$.}
	\label{fig:Fig_IRS_BP_minus20dB_error000}\vspace{-3mm}
\end{figure}

\begin{figure}[!t]\centering\vspace{-0mm}
	\includegraphics[width=0.38\textwidth]{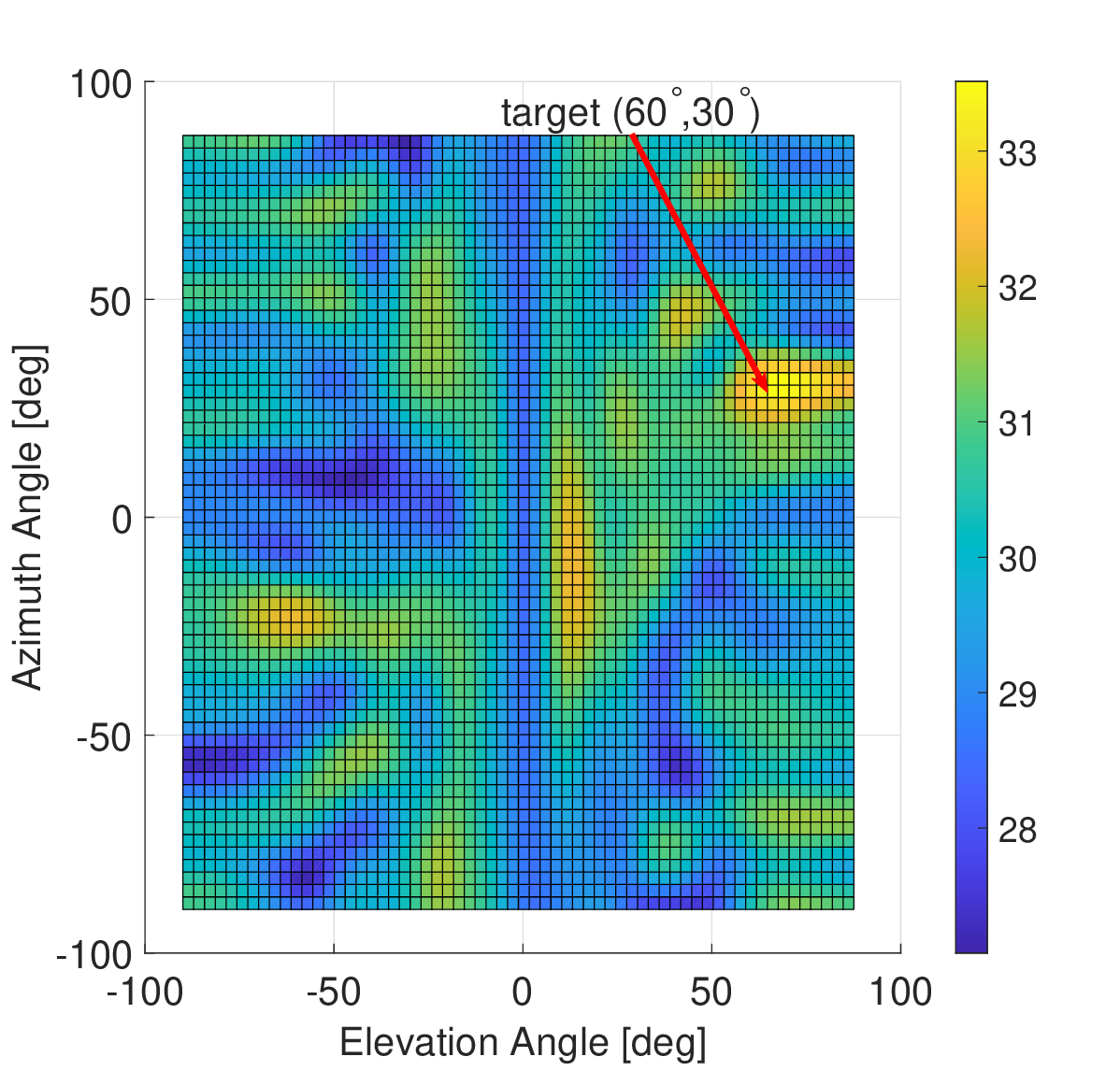}
	\caption{Obtained IRS beampattern; $\omega=0.5$, Rician factors: $-20$\,dB, $\sigma_e^2=-3$\,dB.}
	\label{fig:Fig_IRS_BP_minus20dB_error050}\vspace{-3mm}
\end{figure}

Figs. \ref{fig:Fig_IRS_BP_minus20dB_error000} and \ref{fig:Fig_IRS_BP_minus20dB_error050} illustrate the effect of CSI error on the IRS beampattern where $\omega=0.5$ and the Rician factors of the channels are set as $-20$\,dB. When CSI error is present, the gain of the main lobe towards the target is decreased by $4.5$\,dB by comparing Figs. \ref{fig:Fig_IRS_BP_minus20dB_error000} and \ref{fig:Fig_IRS_BP_minus20dB_error050}. In addition, side lobes appear when  CSI error exists, as Fig. \ref{fig:Fig_IRS_BP_minus20dB_error050} shows. 
When the Rician factors are set as $-20$\,dB,  our proposed qtMM achieves quite stable convergence when $\sigma_e^2$ is smaller than $-3$\,dB (the dimensionless value of $0.50$).   

\vspace{2mm}
\section{CONCLUSIONS}
\label{sec:conclusions}
In this paper, we have considered an IRS-aided DFRC system, where the target to be sensed acts as an eavesdropper. We have proposed an alternating optimization algorithm to jointly design the transmitted  information bearing waveform, the AN, and the IRS parameter matrix, aiming  to maximize the secrecy rate subject to certain non-convex constraints. We have invoked an efficient fractional programming technique and a quadratic transform to convert the fractional objective of secrecy rate into to a more mathematically tractable non-fractional form. Thereby, the design of the waveform and the AN was transformed into a non-fractional quadratic programming problem. The IRS design problem, which contains a challenging fourth order function of the IRS parameter, 
was reduced into a  linear order one by applying  the effective bounding technique, MM, twice. The other quadratic functions of the IRS parameter were downgraded to linear terms via MM.  The transformed problem was solved by solving its dual,  which admits a low-complexity closed-form expression solution. Numerical results are presented to illustrate the convergence properties of the proposed system design method, as well as the secrecy rate and beamforming performance of the designed system. 



  \balance
  \appendices
  \vspace{-0mm}

  \section{Derivation of Lagrangian dual expressions of $R_u$ and $R_{te}$ as (\ref{eqn:Ru_dual}) and (\ref{eqn:Rte_dual})} \label{sec:Lagrangian_dual}
 At first, we show a universal method for converting a  function containing multiple fractional terms, to non-fractional form. Assume an objective is the addition of $I$ terms as {\cite{Shen2018fractional}}
    \begin{eqnarray}
     f(\mathbf x) = \sum_{i=1}^{I} \omega_i \log{\left(1+{A_i(\mathbf x)}/{B_i(\mathbf x)}\right)},
 \end{eqnarray}
 \noindent where $\mathbf x$ is the variable to be optimized, $\omega_i$ is the weight factor for the $i$-th term, $A_i(\mathbf x)$ and $B_i(\mathbf x)$ are polynomials of $\mathbf x$ for $i \in \{1,\cdots,I\}$. Then $f(\mathbf x)$ can be transformed via Lagrangian
dual reformulation as
\addtocounter{equation}{1}
\begin{eqnarray}
 &&f(\mathbf x,\boldsymbol{\gamma}) =\sum_{i=1}^{I} \omega_i \log{\left(1+\gamma_i\right)} - \sum_{i=1}^{I} \omega_i \gamma_i \nonumber \\&&+ \sum_{i=1}^{I} \omega_i (1+\gamma_i) \frac{A_i(\mathbf x)}{A_i(\mathbf x)+B_i(\mathbf x)},  
 \end{eqnarray}
\noindent where $\boldsymbol{\gamma} = [\gamma_1, \cdots, \gamma_i, \cdots \gamma_I]^T$ is an auxiliary variable where $\gamma_i$ can be optimally updated as
     \begin{eqnarray}
         \gamma_i^{\ast} = A_i(\mathbf x) / B_i(\mathbf x), \qquad\forall i = 1,\cdots, I 
     \end{eqnarray}
The optimal $\gamma_i^{\ast}$ above can be easily obtained by taking $\mathbf x$ as a constant and letting $\partial f(\boldsymbol{\gamma})/\partial \gamma_i = 0$. To make the objective be non-fractional, we apply quadratic transform to the third term of the objective as
\addtocounter{equation}{1}
\begin{eqnarray}
\!\!\!\!\!\!\!\!\!\!\!\!\!\!\!\!\!\!\!\!&&f(\mathbf x,\boldsymbol{\gamma}) \rightarrow f(\mathbf x,\boldsymbol{\gamma},\mathbf y) = \sum_{i=1}^{I} \omega_i \log{\left(1+\gamma_i\right)} - \sum_{i=1}^{I} \omega_i \gamma_i \nonumber \\\!\!\!\!\!\!\!\!\!\!\!\!\!\!\!\!\!\!\!\!&&+  \sum_{i=1}^{I} 2 y_i \sqrt{\omega_i (1+\gamma_i)A_i(\mathbf x)} - \sum_{i=1}^{I} y_i^2 (A_i(\mathbf x) + B_i(\mathbf x)), \label{eqn:qt_result}
\end{eqnarray}
\noindent where $\mathbf y = [y_1, \cdots, y_i, \cdots y_I]^T$ is an auxiliary variable and can updated as
     \begin{eqnarray}
         y_i = \sqrt{\omega_i (1+\gamma_i)A_i(\mathbf x)}/(A_i(\mathbf x) + B_i(\mathbf x)),\forall i = 1,\cdots, I
     \end{eqnarray}

By applying the aforementioned method, the fractional function $R_u$ in (\ref{eqn:R_u}) is converted to a non-fractional function as (\ref{eqn:Ru_dual}). Specifically, we replace the variable $\mathbf x$ with $\mathbf w$ and $\mathbf W_n$, and we let $I=1$, $\omega_1=1$, $A_1(\mathbf w, \mathbf W_n)=| \mathbf c_u^T \mathbf w|^2$, $B_1(\mathbf w, \mathbf W_n)=|| \mathbf c_u^T \mathbf W_n||^2 + \sigma_u^2$. Note that  $\gamma_{u}$ and $\alpha_u$  in (\ref{eqn:Ru_dual}) respectively correspond to
 $\gamma_1$ and $y_1$ in (\ref{eqn:qt_result}).
 Similarly, by letting $A_1(\mathbf w, \mathbf W_n) = | \mathbf c_{te}^T \mathbf w|^2$, $B_1(\mathbf w, \mathbf W_n)=|| \mathbf c_{te}^T \mathbf W_n||^2 + \sigma_{te}^2$,
   $R_{te}$ in (\ref{eqn:R_te}) is converted to a non-fractional function as (\ref{eqn:Rte_dual}). 

  \section{Derivation of update of IRS parameter vector in qtMM } \label{sec:phi_update_qtMM}

The Lagrangian multiplier $\rho$, is omitted for the sake of exposition, i.e. $\boldsymbol \phi (\rho)$ is written as $\boldsymbol \phi$ without affecting the results. By invoking
\addtocounter{equation}{2}
  \begin{eqnarray} 
  \!\!\!\!\!\!\!\!\!\!\!\!\!\!\!\!\!\!\!\!\!&&g_0(\boldsymbol \phi) = \boldsymbol \phi^T_t (\mathbf L_1 +\overline{\mathbf L}_1)\boldsymbol \phi^{\ast} + \boldsymbol \phi^T (\mathbf L_1 +\overline{\mathbf L}_1)\boldsymbol \phi^{\ast}_t  \nonumber \\ 
    \!\!\!\!\!\!\!\!\!\!\!\!\!\!\!\!\!\!\!\!\!&&- \boldsymbol \phi^T_t (\mathbf L_1 +\overline{\mathbf L}_1)\boldsymbol \phi_t^{\ast} +\Re{[\boldsymbol{\phi}^T (\boldsymbol \eta-\boldsymbol \mu - \overline{\boldsymbol \mu})]},
    \nonumber \\ \!\!\!\!\!\!\!\!\!\!\!\!\!\!\!\!\!\!\!\!\!&& g_1(\boldsymbol{\phi}) \!=\! 2 \boldsymbol \phi^H_t \!\mathbf L_2 \boldsymbol \phi^{\ast} \!\!\!-\! \boldsymbol \phi^H_t \!\mathbf L_2 \boldsymbol \phi^{\ast}_t \!+\! 2\boldsymbol \phi^T_t \!\mathbf L_3 \boldsymbol \phi   \!-\!\boldsymbol \phi^T_t \!\mathbf L_3 \boldsymbol \phi_t \!-\! \gamma_{\!R,th}',
    \end{eqnarray}
  we have 
  \begin{eqnarray}   \label{eqn:obj_app} \Re{[g_0(\boldsymbol{\phi}) + \rho g_1(\boldsymbol{\phi})]} = \Re{[\boldsymbol \nu^T \boldsymbol \phi^{\ast} + \boldsymbol \kappa^T \boldsymbol \phi + c_L]},
  \end{eqnarray}
\noindent where $\boldsymbol \nu$ and $\boldsymbol \kappa$ are respectively expressed as (\ref{eqn:nu}) and (\ref{eqn:kappa}), and $c_L$ is a constant not relevant to $\boldsymbol \phi$ as
\begin{eqnarray}
    c_L = - \boldsymbol \phi^T_t (\mathbf L_1\!+\!\overline{\mathbf L}_1) \boldsymbol \phi_t^{\ast} \!-\! \rho (\gamma_{R,th} \!+\! \boldsymbol \phi^H_t \!\mathbf L_2 \boldsymbol \phi_t^{\ast} \!+\! \boldsymbol \phi^T_t \!\mathbf L_3 \boldsymbol \phi_t),
\end{eqnarray}
\noindent where $\boldsymbol \phi_t$ is the solved solution of $\boldsymbol \phi$ in $t$-th/previous iteration  and $t$ is the index of outer iteration in Algorithm \ref{alg:alt_opt}. $c_L$ can be dropped safely. The term $\Re{[\boldsymbol \nu^T \boldsymbol \phi^{\ast}]}$ in (\ref{eqn:obj_app}) can be written as
\addtocounter{equation}{1}
\begin{eqnarray}
    \Re{[\boldsymbol \nu^T \boldsymbol \phi^{\ast}]} &=& \Re{ \{ ( \Re{[\boldsymbol{\nu}^T]} + j \Im{[\boldsymbol{\nu}^T]} ) ( \Re{[\boldsymbol{\phi}]} - j \Im{[\boldsymbol{\phi}]} ) \} } \nonumber \\ & =& \Re{[\boldsymbol{\nu}^T]} \Re{[\boldsymbol{\phi}]} + \Im{[\boldsymbol{\nu}^T]}  \Im{[\boldsymbol{\phi}]}.
\end{eqnarray}
In addition, a term $\Re{[\boldsymbol \nu^H \boldsymbol \phi]}$ can be also expressed as
\addtocounter{equation}{1}
\begin{eqnarray}
    \Re{[\boldsymbol \nu^H \boldsymbol \phi]} &=& \Re{ \{ ( \Re{[\boldsymbol{\nu}^T]} - j \Im{[\boldsymbol{\nu}^T]} ) ( \Re{[\boldsymbol{\phi}]} + j \Im{[\boldsymbol{\phi}]} ) \} } \nonumber \\ & =& \Re{[\boldsymbol{\nu}^T]} \Re{[\boldsymbol{\phi}]} + \Im{[\boldsymbol{\nu}^T]}  \Im{[\boldsymbol{\phi}]}.
\end{eqnarray}
Obviously 
\begin{eqnarray}
    \Re{[\boldsymbol \nu^T \boldsymbol \phi^{\ast}]} = \Re{[\boldsymbol \nu^H \boldsymbol \phi]}.
    \end{eqnarray}
Therefore the first two terms in  (\ref{eqn:obj_app}) can be written as
\begin{eqnarray}
    \Re{[\boldsymbol \nu^T \boldsymbol \phi^{\ast} + \boldsymbol \kappa^T \boldsymbol \phi]} = \Re{[\boldsymbol \nu^H \boldsymbol \phi + \boldsymbol \kappa^T \boldsymbol \phi]} = \Re{[(\boldsymbol \nu^H  + \boldsymbol \kappa^T) \boldsymbol \phi]}.
\end{eqnarray}
According to \cite{Shen2019secrecy}, to maximize $\Re{[\mathbf a^H \boldsymbol \phi]}$ where $\mathbf a \in \mathbb C^{N \times 1}$,  the phase of the $n$-th element of $\mathbf a$ and that of $\boldsymbol \phi$ should be equal $\forall n \in \{1,\cdots,N\}$, i.e.
\addtocounter{equation}{1}
\begin{eqnarray}
   \boldsymbol \phi &=& [ \exp{( j \arg[a_1] )}, \cdots, \exp{( j \arg[a_N] )} ]^T \nonumber \\&=& \exp{( j \arg[\mathbf a] )}. 
\end{eqnarray}
Likewise, the $\boldsymbol \phi$ that maximize $\Re{[(\boldsymbol \nu^H  + \boldsymbol \kappa^T) \boldsymbol \phi]}$ is
\addtocounter{equation}{0}
\begin{eqnarray} 
    \boldsymbol{\phi} = \exp{[j \arg( [\boldsymbol \nu^H  + \boldsymbol \kappa^T]^H )]} =  \exp{[j \arg(\boldsymbol \nu + \boldsymbol \kappa^{\ast})]}.
\end{eqnarray}


\vspace{12pt}
\balance

\bibliographystyle{IEEEtran}
\bibliography{IEEEabrv,References,Ref2}

\begin{thebibliography}{10}
\providecommand{\url}[1]{#1}
\csname url@samestyle\endcsname
\providecommand{\newblock}{\relax}
\providecommand{\bibinfo}[2]{#2}
\providecommand{\BIBentrySTDinterwordspacing}{\spaceskip=0pt\relax}
\providecommand{\BIBentryALTinterwordstretchfactor}{4}
\providecommand{\BIBentryALTinterwordspacing}{\spaceskip=\fontdimen2\font plus
\BIBentryALTinterwordstretchfactor\fontdimen3\font minus
  \fontdimen4\font\relax}
\providecommand{\BIBforeignlanguage}[2]{{%
\expandafter\ifx\csname l@#1\endcsname\relax
\typeout{** WARNING: IEEEtran.bst: No hyphenation pattern has been}%
\typeout{** loaded for the language `#1'. Using the pattern for}%
\typeout{** the default language instead.}%
\else
\language=\csname l@#1\endcsname
\fi
#2}}
\providecommand{\BIBdecl}{\relax}
\BIBdecl

\bibitem{Li2023anIRS}
Y.-K. Li and A.~Petropulu, ``An {IRS}-assisted secure dual-function
  radar-communication system,'' \emph{arXiv:2310.00555}, 2023.

\bibitem{Chen2020performance}
X.~Chen, Z.~Feng, Z.~Wei, F.~Gao, and X.~Yuan, ``Performance of joint
  sensing-communication cooperative sensing {UAV} network,'' \emph{{IEEE}
  Trans. Veh. Technol.}, vol.~69, no.~12, pp. 15\,545--15\,556, 2020.

\bibitem{Feng2020joint}
Z.~Feng, Z.~Fang, Z.~Wei, X.~Chen, Z.~Quan, and D.~Ji, ``Joint radar and
  communication: A survey,'' \emph{China Commun.}, vol.~17, no.~1, pp. 1--27,
  2020.

\bibitem{Zhang2021anoverview}
J.~A. Zhang, F.~Liu, C.~Masouros, R.~W. Heath, Z.~Feng, L.~Zheng, and
  A.~Petropulu, ``An overview of signal processing techniques for joint
  communication and radar sensing,'' \emph{IEEE Journal of Selected Topics in
  Signal Processing}, vol.~15, no.~6, pp. 1295--1315, 2021.

\bibitem{Sturm2011waveform}
C.~Sturm and W.~Wiesbeck, ``Waveform design and signal processing aspects for
  fusion of wireless communications and radar sensing,'' \emph{Proc. {IEEE}},
  vol.~99, no.~7, pp. 1236--1259, 2011.

\bibitem{Mu2023UAV}
J.~Mu, R.~Zhang, Y.~Cui, N.~Gao, and X.~Jing, ``{UAV} meets integrated sensing
  and communication: Challenges and future directions,'' \emph{{IEEE} Commun.
  Mag.}, vol.~61, no.~5, pp. 62--67, 2023.

\bibitem{Ma2023integrated}
T.~Ma, Y.~Xiao, X.~Lei, and M.~Xiao, ``Integrated sensing and communication for
  wireless extended reality ({XR}) with reconfigurable intelligent surface,''
  \emph{{IEEE} J. Sel. Topics Signal Process.}, vol.~17, no.~5, pp. 980--994,
  2023.

\bibitem{Zhao2022joint}
X.~Zhao and Y.-J.~A. Zhang, ``Joint beamforming and scheduling for integrated
  sensing and communication systems in {URLLC},'' in \emph{GLOBECOM 2022 - 2022
  IEEE Global Communications Conference}, 2022, pp. 3611--3616.

\bibitem{Zhong2022empowering}
Y.~Zhong, T.~Bi, J.~Wang, J.~Zeng, Y.~Huang, T.~Jiang, Q.~Wu, and S.~Wu,
  ``Empowering the {V2X} network by integrated sensing and communications:
  Background, design, advances, and opportunities,'' \emph{{IEEE} Netw.},
  vol.~36, no.~4, pp. 54--60, 2022.

\bibitem{Liu2020joint}
F.~Liu, C.~Masouros, A.~P. Petropulu, H.~Griffiths, and L.~Hanzo, ``Joint radar
  and communication design: Applications, state-of-the-art, and the road
  ahead,'' \emph{{IEEE} Trans. Commun.}, vol.~68, no.~6, pp. 3834--3862, 2020.

\bibitem{Hassanien2019dual}
A.~Hassanien, M.~G. Amin, E.~Aboutanios, and B.~Himed, ``Dual-function radar
  communication systems: A solution to the spectrum congestion problem,''
  \emph{{IEEE} Signal Process. Mag.}, vol.~36, no.~5, pp. 115--126, 2019.

\bibitem{Ma2020joint}
D.~Ma, N.~Shlezinger, T.~Huang, Y.~Liu, and Y.~C. Eldar, ``Joint
  radar-communication strategies for autonomous vehicles: Combining two key
  automotive technologies,'' \emph{{IEEE} Signal Process. Mag.}, vol.~37,
  no.~4, pp. 85--97, 2020.

\bibitem{Wyner1975thewire}
A.~D. Wyner, ``The wire-tap channel,'' \emph{The Bell System Technical
  Journal}, vol.~54, no.~8, pp. 1355--1387, 1975.

\bibitem{Fakoorian2011solutions}
S.~A.~A. Fakoorian and A.~L. Swindlehurst, ``Solutions for the {MIMO} gaussian
  wiretap channel with a cooperative jammer,'' \emph{{IEEE} Trans. Signal
  Process.}, vol.~59, no.~10, pp. 5013--5022, 2011.

\bibitem{Dong2010improving}
L.~Dong, Z.~Han, A.~P. Petropulu, and H.~V. Poor, ``Improving wireless physical
  layer security via cooperating relays,'' \emph{{IEEE} Trans. Signal
  Process.}, vol.~58, no.~3, pp. 1875--1888, 2010.

\bibitem{Zheng2011optimal}
G.~Zheng, L.-C. Choo, and K.-K. Wong, ``Optimal cooperative jamming to enhance
  physical layer security using relays,'' \emph{{IEEE} Trans. Signal Process.},
  vol.~59, no.~3, pp. 1317--1322, 2011.

\bibitem{Li2011oncooperative}
J.~Li, A.~P. Petropulu, and S.~Weber, ``On cooperative relaying schemes for
  wireless physical layer security,'' \emph{{IEEE} Trans. Signal Process.},
  vol.~59, no.~10, pp. 4985--4997, 2011.

\bibitem{Zheng2013improving}
G.~Zheng, I.~Krikidis, J.~Li, A.~P. Petropulu, and B.~Ottersten, ``Improving
  physical layer secrecy using full-duplex jamming receivers,'' \emph{{IEEE}
  Trans. Signal Process.}, vol.~61, no.~20, pp. 4962--4974, 2013.

\bibitem{Khisti2010secure}
A.~Khisti and G.~W. Wornell, ``Secure transmission with multiple antennas {I}:
  The {MISOME} wiretap channel,'' \emph{{IEEE} Trans. Inf. Theory}, vol.~56,
  no.~7, pp. 3088--3104, 2010.

\bibitem{Goel2008guaranteeing}
S.~Goel and R.~Negi, ``Guaranteeing secrecy using artificial noise,''
  \emph{{IEEE} Trans. Wireless Commun.}, vol.~7, no.~6, pp. 2180--2189, 2008.

\bibitem{Su2021secure}
N.~Su, F.~Liu, C.~Masouros, T.~Ratnarajah, and A.~Petropulu, ``Secure
  dual-functional radar-communication transmission: Hardware-efficient
  design,'' in \emph{2021 55th Asilomar Conference on Signals, Systems, and
  Computers}, 2021, pp. 629--633.

\bibitem{Su2023sensing}
N.~Su, F.~Liu, and C.~Masouros, ``Sensing-assisted eavesdropper estimation: An
  {ISAC} breakthrough in physical layer security,'' \emph{arXiv:2210.08286},
  2023.

\bibitem{Wei2022multiple}
T.~Wei, L.~Wu, K.~V. Mishra, and M.~R.~B. Shankar, ``Multiple {IRS}-assisted
  wideband dual-function radar-communication,'' in \emph{2022 2nd IEEE
  International Symposium on Joint Communications \& Sensing (JC\&S)}, 2022,
  pp. 1--5.

\bibitem{Jiang2021intelligent}
Z.-M. Jiang, M.~Rihan, P.~Zhang, L.~Huang, Q.~Deng, J.~Zhang, and E.~M.
  Mohamed, ``Intelligent reflecting surface aided dual-function radar and
  communication system,'' \emph{{IEEE} Syst. J.}, pp. 1--12, 2021.

\bibitem{Li2022dual}
Y.~Li and A.~Petropulu, ``Dual-function radar-communication system aided by
  intelligent reflecting surfaces,'' in \emph{2022 IEEE 12th Sensor Array and
  Multichannel Signal Processing Workshop (SAM)}, 2022, pp. 126--130.

\bibitem{Li2023minorization}
Y.-K. Li and A.~Petropulu, ``Minorization-based low-complexity design for
  {IRS}-aided {ISAC} systems,'' in \emph{2023 IEEE Radar Conference
  (RadarConf23)}, 2023, pp. 1--6.

\bibitem{Liu2022joint}
R.~Liu, M.~Li, Y.~Liu, Q.~Wu, and Q.~Liu, ``Joint transmit waveform and passive
  beamforming design for {RIS}-aided {DFRC} systems,'' \emph{{IEEE} J. Sel.
  Topics Signal Process.}, vol.~16, no.~5, pp. 995--1010, 2022.

\bibitem{Li2023intelligent}
Y.-K. Li and A.~Petropulu, ``Intelligent reflecting surface-assisted
  dual-function radar-communication system,'' \emph{{IEEE} Access}, pp. 1--1,
  2023.

\bibitem{Sun2017majorization}
Y.~Sun, P.~Babu, and D.~P. Palomar, ``Majorization-minimization algorithms in
  signal processing, communications, and machine learning,'' \emph{{IEEE}
  Trans. Signal Process.}, vol.~65, no.~3, pp. 794--816, 2017.

\bibitem{Shen2018fractional}
K.~Shen and W.~Yu, ``Fractional programming for communication systems—part
  {I}: Power control and beamforming,'' \emph{{IEEE} Trans. Signal Process.},
  vol.~66, no.~10, pp. 2616--2630, 2018.

\bibitem{Najafi2021physics}
M.~Najafi, V.~Jamali, R.~Schober, and H.~V. Poor, ``Physics-based modeling and
  scalable optimization of large intelligent reflecting surfaces,''
  \emph{{IEEE} Trans. Commun.}, vol.~69, no.~4, pp. 2673--2691, 2021.

\bibitem{Li2023efficient}
Y.-K. Li and A.~Petropulu, ``Efficient beamforming designs for {IRS}-aided
  {DFRC} systems,'' in \emph{2023 24th International Conference on Digital
  Signal Processing (DSP)}, 2023, pp. 1--5.

\bibitem{He2022qcqp}
X.~He and J.~Wang, ``{QCQP} with extra constant modulus constraints: Theory and
  application to {SINR} constrained mm{w}ave hybrid beamforming,'' \emph{{IEEE}
  Trans. Signal Process.}, vol.~70, pp. 5237--5250, 2022.

\bibitem{Wang2023star}
C.~Wang, C.-C. Wang, Z.~Li, D.~W.~K. Ng, K.-K. Wong, N.~Al-Dhahir, and
  D.~Niyato, ``{STAR}-{RIS}-enabled secure dual-functional
  radar-communications: Joint waveform and reflective beamforming
  optimization,'' \emph{{IEEE} Trans. Inf. Forensics Security}, vol.~18, pp.
  4577--4592, 2023.

\bibitem{Hua2023secure}
M.~Hua, Q.~Wu, W.~Chen, O.~A. Dobre, and A.~Lee~Swindlehurst, ``Secure
  intelligent reflecting surface aided integrated sensing and communication,''
  \emph{{IEEE} Trans. Wireless Commun.}, pp. 1--1, 2023.

\bibitem{Chu2023joint}
J.~Chu, Z.~Lu, R.~Liu, M.~Li, and Q.~Liu, ``Joint beamforming and reflection
  design for secure {RIS-ISAC} systems,'' \emph{{IEEE} Trans. Veh. Technol.},
  pp. 1--5, 2023.

\bibitem{Zhang2023irs}
H.~Zhang and J.~Zheng, ``{IRS}-assisted secure radar communication systems with
  malicious target,'' \emph{{IEEE} Trans. Veh. Technol.}, pp. 1--14, 2023.

\bibitem{Sweta2023efficient}
S.~Sweta, A.~Dubey, and V.-L. Nguyen, ``Efficient {IRS}-aided rate
  optimizations for dual functional radar and communications,'' \emph{{IEEE}
  Commun. Lett.}, pp. 1--1, 2023.

\bibitem{Zhao2023joint}
H.~Zhao, F.~Wu, W.~Xia, Y.~Zhang, Y.~Ni, and H.~Zhu, ``Joint beamforming design
  for {RIS}-aided secure integrated sensing and communication systems,''
  \emph{{IEEE} Commun. Lett.}, vol.~27, no.~11, pp. 2943--2947, 2023.

\bibitem{Salem2022active}
A.~A. Salem, M.~H. Ismail, and A.~S. Ibrahim, ``Active reconfigurable
  intelligent surface-assisted {MISO} integrated sensing and communication
  systems for secure operation,'' \emph{{IEEE} Trans. Veh. Technol.}, pp.
  1--13, 2022.

\bibitem{Mishra2022optm3sec}
K.~V. Mishra, A.~Chattopadhyay, S.~S. Acharjee, and A.~P. Petropulu,
  ``Optm3sec: Optimizing multicast {IRS}-aided multiantenna {DFRC} secrecy
  channel with multiple eavesdroppers,'' in \emph{ICASSP 2022 - 2022 IEEE
  International Conference on Acoustics, Speech and Signal Processing
  (ICASSP)}, 2022, pp. 9037--9041.

\bibitem{Dinkelbach1967onnonlinear}
W.~Dinkelbach, ``On nonlinear fractional programming,'' \emph{Management
  Science}, vol.~13, no.~7, pp. 492--498, 1967.

\bibitem{Evmorfos2023actor}
S.~Evmorfos, A.~P. Petropulu, and H.~V. Poor, ``Actor-critic methods for {IRS}
  design in correlated channel environments: A closer look into the neural
  tangent kernel of the critic,'' \emph{{IEEE} Trans. Signal Process.},
  vol.~71, pp. 4029--4044, 2023.

\bibitem{Liu2023drl}
Q.~Liu, Y.~Zhu, M.~Li, R.~Liu, Y.~Liu, and Z.~Lu, ``{DRL}-based secrecy rate
  optimization for {RIS}-assisted secure {ISAC} systems,'' \emph{{IEEE} Trans.
  Veh. Technol.}, pp. 1--5, 2023.

\bibitem{Jiang2023sensing}
S.~Jiang and A.~Alkhateeb, ``Sensing aided {OTFS} massive {MIMO} systems:
  Compressive channel estimation,'' in \emph{2023 IEEE International Conference
  on Communications Workshops (ICC Workshops)}, 2023, pp. 794--799.

\bibitem{Silvia2023enhanced}
S.~Mura, M.~Mizmizi, U.~Spagnolini, and A.~Petropulu, ``Enhanced channel
  estimation in mm-wave {MIMO} systems leveraging integrated communication and
  sensing,'' \emph{arXiv:2309.14875}, 2023.

\bibitem{Li2017_coexistence}
B.~Li and A.~P. Petropulu, ``Joint transmit designs for coexistence of {MIMO}
  wireless communications and sparse sensing radars in clutter,'' \emph{{IEEE}
  Trans. Aerosp. Electron. Syst.}, vol.~53, no.~6, pp. 2846--2864, 2017.

\bibitem{cvx}
M.~Grant and S.~Boyd, ``{CVX}: Matlab software for disciplined convex
  programming, version 2.1,'' \url{http://cvxr.com/cvx}, Mar. 2014.

\bibitem{Shen2019secrecy}
H.~Shen, W.~Xu, S.~Gong, Z.~He, and C.~Zhao, ``Secrecy rate maximization for
  intelligent reflecting surface assisted multi-antenna communications,''
  \emph{{IEEE} Commun. Lett.}, vol.~23, no.~9, pp. 1488--1492, 2019.

\end{thebibliography}

\end{document}